%% file: main.tex
\documentclass[lettersize,journal]{IEEEtran}
\usepackage{amsmath,amsfonts}
\usepackage{array}
\usepackage{textcomp}
\usepackage{stfloats}
\usepackage{url}
\usepackage{verbatim}
\usepackage{graphicx}
\usepackage{cite}
\usepackage[free-standing-units,per-mode=repeated-symbol,mode=text,detect-weight=true,detect-family=true]{siunitx}
\usepackage{glossaries}
\usepackage{enumitem}
\usepackage{nicefrac}
\usepackage{todonotes}
\usepackage{xcolor}
\usepackage{menukeys}
\usepackage[ruled,vlined]{algorithm2e}
\usepackage{multirow}
\usepackage[flushleft]{threeparttable}
\usepackage{circledsteps}
\usepackage{threeparttable}
\usepackage{microtype}
\usepackage{hyperref}
\usepackage[capitalise,nameinlink]{cleveref}
\usepackage{booktabs}
\usepackage{multirow}
\usepackage{tabularx}
\usepackage{graphicx}
\usepackage{caption}
\usepackage{xspace}
\usepackage{subcaption}
\usepackage[normalem]{ulem}

\DeclareSIUnit\core{core}
\DeclareSIUnit\request{req}
\DeclareSIUnit\cycle{cycle}
\DeclareSIUnit\erlang{E}
\DeclareSIUnit\flop{FLOP}
\DeclareSIUnit\flops{FLOPS}
\DeclareSIUnit\gflops{GFLOPS}
\DeclareSIUnit\tflops{TFLOPS}
\DeclareSIUnit\gate{GE}
\DeclareSIUnit\ge{GE}
\DeclareSIUnit\op{OP}
\DeclareSIUnit\ops{OPS}
\DeclareSIUnit\bps{bps}
\DeclareSIUnit\Bps{Bps}
\DeclareSIUnit\ipc{IPC}
\DeclareSIUnit\token{Token}
\DeclareSIUnit[number-unit-product = ]\percent{\%}

\input{acr/acronyms}

\definecolor{done}{rgb}{0.0, 0.5, 0.0}
\definecolor{wip}{rgb}{0.8, 0.5, 0.21}
\definecolor{todo}{rgb}{0.52, 0.52, 0.51}
\definecolor{question}{rgb}{0.36, 0.75, 0.92}
\definecolor{green}{rgb}{0.51, 0.7, 0.4}
\definecolor{red}{rgb}{0.8, 0, 0.4}
\definecolor{lp}{rgb}{0.39, 0.46, 0.53}
\definecolor{fa2}{rgb}{0.43, 0.53, 0.39}
\definecolor{sa}{rgb}{0.46, 0.38, 0.54}
\definecolor{PULPOrange}{HTML}{F29545}
\definecolor{mossgreen}{HTML}{5FAD56}

\newcommand\softmax{Softmax\xspace}
\newcommand{\riscv}{RISC-V\xspace}
\newcommand\snitch{Snitch\xspace}
\newcommand\nvidia{Nvidia\xspace}
\newcommand\fpd{\texttt{\si{FP64}}\xspace}
\newcommand\fps{\texttt{\si{FP32}}\xspace}
\newcommand\fph{\texttt{\si{FP16}}\xspace}
\newcommand\bfh{\texttt{\si{BF16}}\xspace}
\newcommand\fpb{\texttt{\si{FP8}}\xspace}
\newcommand\fpba{\texttt{\si{FP8ALT}}\xspace}

\newcommand\gelu{GELU\xspace}
\newcommand\fa{FlashAttention-2\xspace}
\newcommand\x{$\times$\xspace}
\newcommand{\citet}{[x]}
\newcommand\gpt{\gls{gpt}\xspace}
\newcommand\vit{\gls{vit}\xspace}

\newcommand\etal{et\penalty50\ al.}

\newcommand\numberincircle[1]{\Circled[inner color=white, outer color=white, fill color=PULPOrange]{#1}}

\newcommand\blackcircle[1]{\Circled[inner color=white, outer color=white, fill color=black]{#1}}

\newcommand\trafocircle[1]{\Circled[inner color=white, outer color=white, fill color=mossgreen]{#1}}

\providecommand{\comment}[1]{\textcolor{black}{}}

\setlist[description]{leftmargin=0cm,labelindent=0cm}

\begin{document}
\bstctlcite{IEEEexample:BSTcontrol}

\title{Optimizing Foundation Model Inference on a Many-tiny-core Open-source RISC-V Platform}
\author{Viviane Potocnik, Luca Colagrande, Tim Fischer, Luca Bertaccini, Daniele Jahier Pagliari, Alessio Burrello, Luca Benini
\thanks{Viviane Potocnik, Luca Colagrande, Tim Fischer, Luca Bertaccini, L. Benini are with the Integrated Systems Laboratory (IIS) of ETH Z\"urich, ETZ, Gloriastrasse 35, 8092 Z\"urich, Switzerland (e-mail: name.surname@iis.ee.ethz.ch).}
\thanks{A. Burrello, L. Benini are with the Department of Electrical, Electronic and Information Engineering, University of Bologna, 40136 Bologna, Italy.
E-mail: firstname.firstsurname@unibo.it}
\thanks{A. Burrello, D. Jahier Pagliari are with the Department of Control and Computer Engineering, Politecnico di Torino, 10129, Turin, Italy. E-mail: name.surname@polito.it}
\thanks{Manuscript received xxx 2024; revised xxx 2024.}}

\markboth{Journal of \LaTeX\ Class Files,~Vol.~14, No.~8, August~2021}%
{Shell \MakeLowercase{\textit{et al.}}: A Sample Article Using IEEEtran.cls for IEEE Journals}


\maketitle
\input{Sections/00_Abstract}


\input{Sections/01_Introduction}
\input{Sections/02_Background}

\input{Sections/03_RelatedWork}

\input{Sections/04_HW_platform}
\input{Sections/05_fm_library}
\input{Sections/06_Results}
\input{Sections/07_Conclusions}

\bibliographystyle{IEEEtran}
\bibliography{\jobname, occamy_llm}

\vspace{-3em}
\begin{IEEEbiography}[{\includegraphics[width=1in,height=1.25in,clip,keepaspectratio]{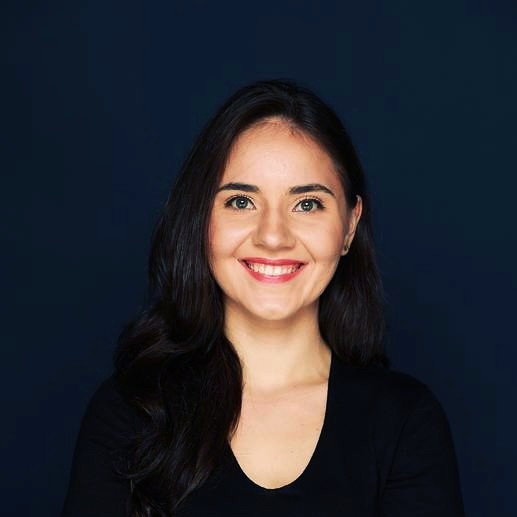}}]{Viviane Potocnik}
received her BSc and MSc degree in Electrical
Engineering and Information Technology from ETH Zurich in 2020 and 2022.
She is currently pursuing a PhD in the Digital Circuits and Systems group of Prof. Benini.
Her research focuses on heterogeneous architectures for energy-efficient multi-modal AI fusion and the exploration of innovative data representation strategies to enhance the computational efficiency and adaptability on devices at the extreme edge, ranging from high-performance to resource-constrained environments.
\end{IEEEbiography}
\vspace{-3em}
\begin{IEEEbiography}[{\includegraphics[width=1in,height=1.25in,clip,keepaspectratio]{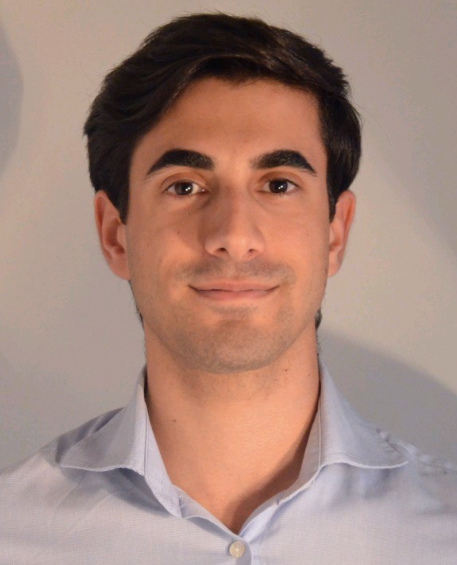}}]{Luca Colagrande}
received his BSc degree from Politecnico di Milano in 2018 and his MSc degree from ETH Zurich in 2020.
He is currently pursuing a PhD in the Digital Circuits and Systems group of Prof. Benini.
His research focuses on the co-design of energy-efficient general-purpose manycore accelerators for machine learning and high-performance computing applications.
\end{IEEEbiography}
\begin{IEEEbiography}[{\includegraphics[width=1in,height=1.25in,clip,keepaspectratio]{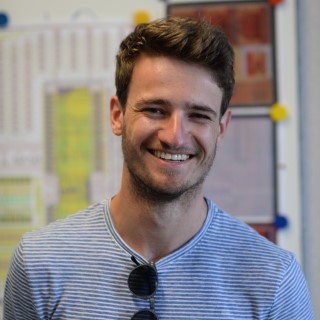}}]{Tim Fischer}
received his BSc and MSc in ``Electrical Engineering and Information Technology'' from the Swiss Federal Institute of Technology Zurich (ETHZ), Switzerland, in 2018 and 2021, respectively. He is currently pursuing a Ph.D. degree at ETH Zurich in the Digital Circuits and Systems group led by Prof. Luca Benini. His research interests include scalable and energy-efficient interconnects for both on-chip and off-chip communication.
\end{IEEEbiography}
\vspace{-3em}
\begin{IEEEbiography}[{\includegraphics[width=1.05in,height=1.25in,clip,keepaspectratio]{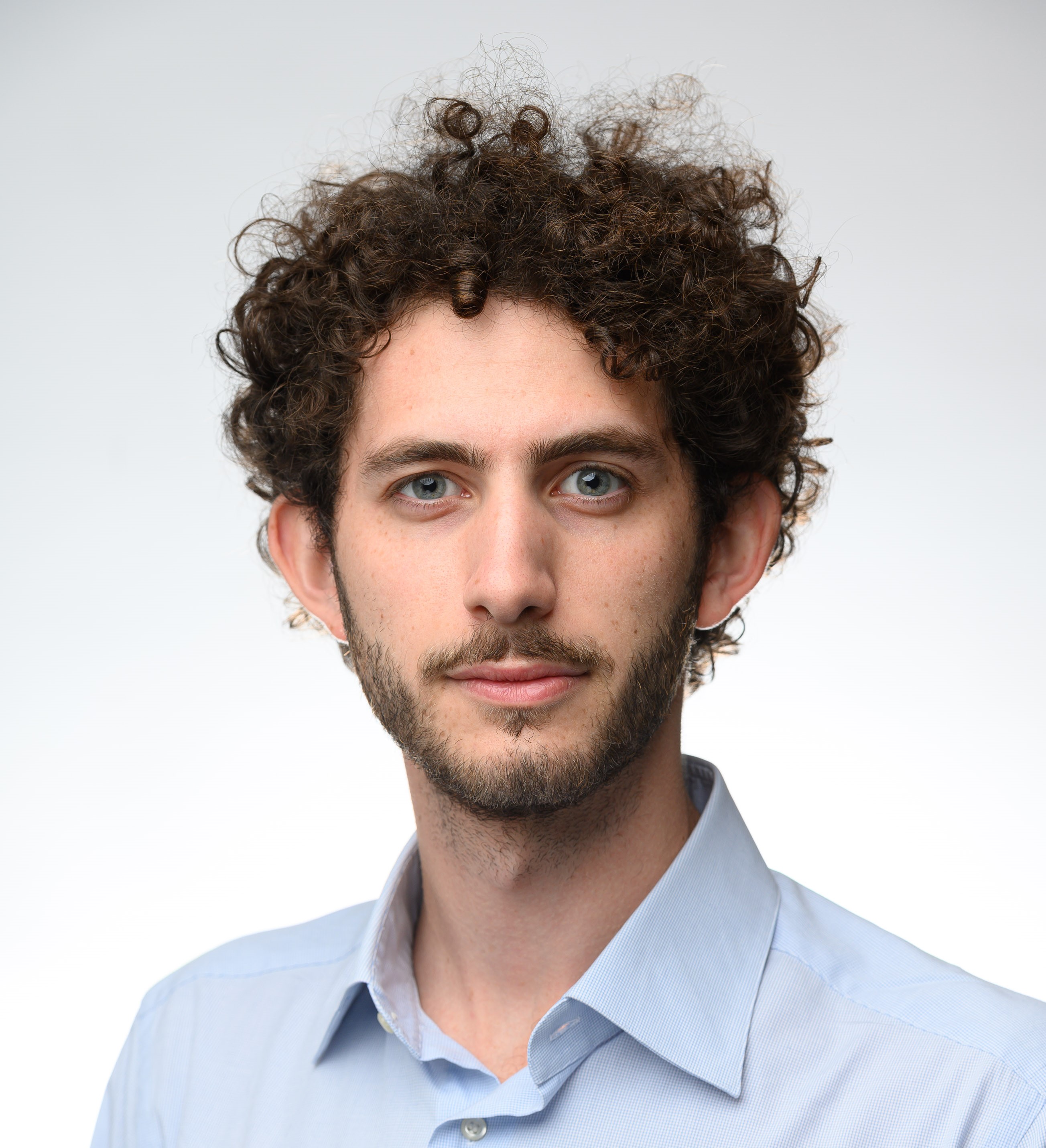}}]{Luca Bertaccini}
received the M.Sc. degree in Electronic Engineering from the University of Bologna in 2020. He is currently pursuing a Ph.D. degree at ETH Zurich in the Digital Circuits and Systems group led by Prof. Luca Benini. His research interests include heterogeneous systems-on-chip, energy-efficient hardware accelerators, computer arithmetic, and transprecision computing.
\end{IEEEbiography}
\vspace{-3em}
\begin{IEEEbiography}[{\includegraphics[width=1in,height=1.25in,clip,keepaspectratio]{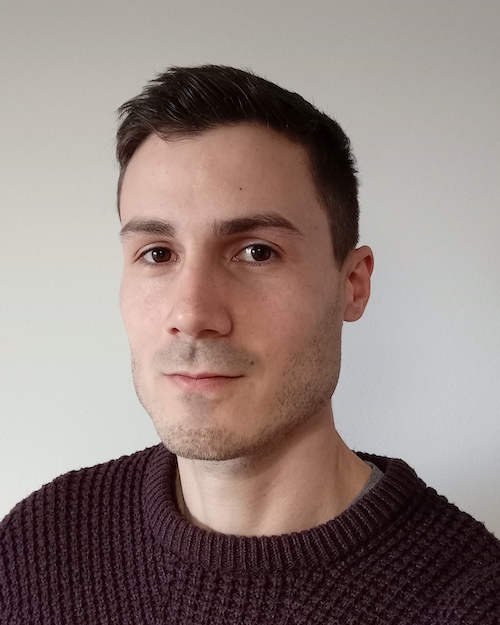}}]{Daniele Jahier Pagliari} received the M.Sc. and Ph.D. degrees in computer engineering from the Politecnico di Torino, Turin, Italy, in 2014 and 2018, respectively. He is currently an Assistant Professor with the Politecnico di Torino. His research interests are in the computer-aided design and optimization of digital circuits and systems, with a particular focus on energy-efficiency aspects and on emerging applications, such as machine learning at the edge.
\end{IEEEbiography}
\vspace{-3em}
\begin{IEEEbiography}[{\includegraphics[width=1in,height=1.25in,clip,keepaspectratio]{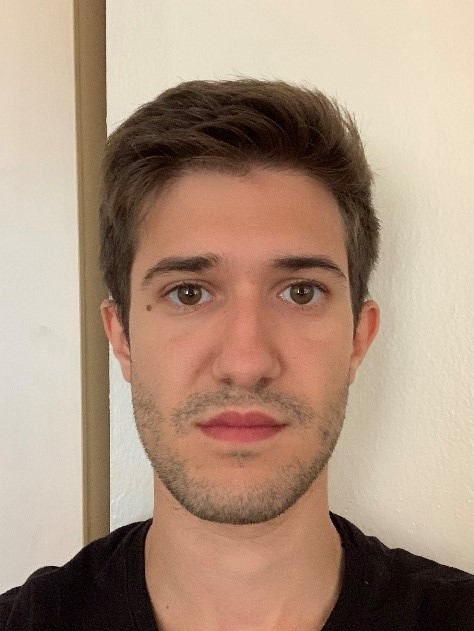}}]{Alessio Burrello}
is currently a research assistant at Politecnico di Torino. He received his Ph.D. degrees in Electronic Engineering at the University of Bologna in 2023. His research interests include parallel programming models for embedded systems, machine and deep learning, hardware-oriented deep learning, and code optimization for multi-core systems. He has published over 90 papers in peer-reviewed international journals and conferences. 
\end{IEEEbiography}
\vspace{-3em}
\begin{IEEEbiography}[{\includegraphics[width=1in,height=1.25in,clip,keepaspectratio]{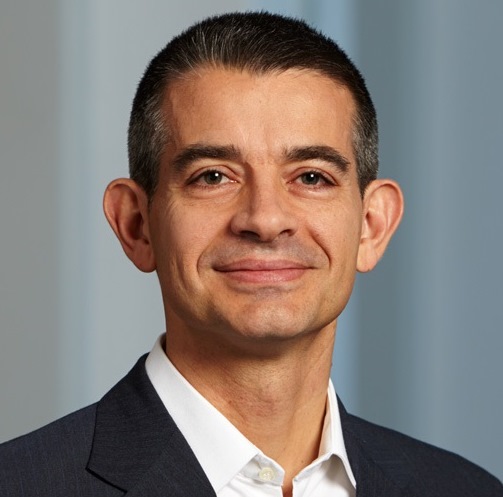}}]{Luca Benini} holds the chair of digital Circuits and systems at ETHZ and is Full Professor at the Universita di Bologna. He received his Ph.D. from Stanford University.  Dr. Benini’s research interests are in energy-efficient parallel computing systems, smart sensing micro-systems, and machine learning hardware. He is a Fellow of the IEEE, of the ACM, and a member of the Academia Europaea.
\end{IEEEbiography}

\end{document}

%% file: acr/acronyms.tex
\newacronym{abi}{ABI}{application binary interface}
\newacronym{ace}{ACE}{AXI Coherent Extensions}
\newacronym{alu}{ALU}{arithmetic logic unit}
\newacronym{amba}{AMBA}{Advanced Microcontroller Bus Architecture}
\newacronym{amo}{AMO}{atomic memory operation}
\newacronym{aot}{AOT}{ahead-of-time}
\newacronym{ai}{AI}{artificial intelligence}
\newacronym{apb}{APB}{Advanced Peripheral Bus}
\newacronym{api}{API}{application programming interface}
\newacronym{asic}{ASIC}{application-specific integrated circuit}
\newacronym{axi}{AXI}{Advanced eXtensible Interface}
\newacronym{bfs}{BFS}{breadth-first search}
\newacronym{blas}{BLAS}{Basic Linear Algebra Subprograms}
\newacronym{cas}{CAS}{compare-and-swap}
\newacronym{cim}{CIM}{Compute-In-Memory}
\newacronym{csp}{CSP}{Cross-layer Sparse Prediction}
\newacronym{cmos}{CMOS}{complementary metal-oxide-semiconductor}
\newacronym{cnn}{CNN}{convolutional neural network}
\newacronym{cpu}{CPU}{central processing unit}
\newacronym{csr}{CSR}{control and status register}
\newacronym{cv}{CV}{computer vision}
\newacronym{dbt}{DBT}{dynamic binary translation}
\newacronym{dct}{DCT}{discrete cosine transform}
\newacronym{dlp}{DLP}{data level parallelism}
\newacronym{dma}{DMA}{direct memory access}
\newacronym{dnn}{DNN}{deep neural network}
\newacronym{dram}{DRAM}{dynamic random-access memory}
\newacronym{dsl}{DSL}{domain-specific language}
\newacronym{dsp}{DSP}{digital signal processing}
\newacronym{elf}{ELF}{Executable and Linkable Format}
\newacronym{fdsoi}{FD-SOI}{fully depleted silicon-on-insulator}
\newacronym{fpga}{FPGA}{field-programmable gate array}
\newacronym{fp}{FP}{floating-point}
\newacronym{fpu}{FPU}{floating-point unit}
\newacronym{frep}{FREP}{FP repetition}
\newacronym{fma}{FMA}{fused multiply-accumulate}
\newacronym{gelu}{GELU}{Gaussian Error Linear Unit}
\newacronym{fmac}{FMAC}{fused multiply-accumulate}
\newacronym{q}{Q}{Queries}
\newacronym{k}{K}{Keys}
\newacronym{v}{V}{Values}
\newacronym{e}{E}{Embedding}
\newacronym{p}{P}{Projection}
\newacronym{h}{H}{head}
\newacronym{gemm}{GEMM}{General Matrix Multiplication}
\newacronym[longplural={general-purpose \acrlongpl{gpu}}]{gpgpu}{GPGPU}{general-purpose \gls{gpu}}
\newacronym{gpt}{GPT}{Generative Pretrained Transformer}
\newacronym{gpu}{GPU}{graphics processing unit}
\newacronym{gl}{GL}{graph learning}
\newacronym{hw}{hardware}{hardware}
\newacronym{hart}{hart}{hardware thread}
\newacronym{hbm}{HBM}{High Bandwidth Memory}
\newacronym{hdl}{HDL}{hardware description language}
\newacronym{hdpe}{HDPE}{Heterogeneous Dataflow Processing Element}
\newacronym{hdsc}{HDSC}{Hybrid Multiplication-Accumulation Unit}
\newacronym{hero}{HERO}{Heterogeneous Embedded Research Platform}
\newacronym{hmau}{HMAU}{Hybrid Multiplication-Accumulation Unit}
\newacronym{hpc}{HPC}{high-performance computing}
\newacronym{ilp}{ILP}{instruction level parallelism}
\newacronym{idol}{I\$}{instruction cache}
\newacronym{ipc}{IPC}{instructions per cycle}
\newacronym{iot}{IoT}{Internet of Things}
\newacronym{ipu}{IPU}{integer processing unit}
\newacronym{ir}{IR}{intermediate representation}
\newacronym{isa}{ISA}{instruction set architecture}
\newacronym{issr}{ISSR}{indirection stream semantic register}
\newacronym{iwgu}{IWGU}{Implicit Weight Generation Unit }
\newacronym{jit}{JIT}{just-in-time}
\newacronym[firstplural=Large Language Models]{llm}{LLM}{Large Language Model}
\newacronym{lare}{LARE}{Local-Attention-Reusable Engine}
\newacronym[firstplural=Small Language Models]{slm}{SLM}{Small Language Model}
\newacronym[firstplural=Language Models]{lm}{LM}{Language Model}
\newacronym{llc}{LLC}{last-level cache}
\newacronym{lrsc}{LRSC}{load-reserved/store-conditional}
\newacronym{lpu}{LPU}{Language Processing Unit}
\newacronym{lr}{LR}{load-reserved}
\newacronym{lsu}{LSU}{load-store unit}
\newacronym{mac}{MAC}{multiply–accumulate}
\newacronym{mimd}{MIMD}{multiple instruction, multiple data}
\newacronym{mp}{MP}{Mixed-Precision}
\newacronym{mhsa}{MHSA}{Multi-Head Self-Attention}
\newacronym{mha}{MHA}{Multi-Head Attention}
\newacronym{gqa}{GQA}{Grouped-Query Attention}
\newacronym{ml}{ML}{Machine Learning}
\newacronym{mlp}{MLP}{multi-layer perceptron}
\newacronym{mmu}{MMU}{memory management unit}
\newacronym{nlp}{NLP}{natural language processing}
\newacronym[longplural={networks-on-chip}]{noc}{NoC}{Network-on-Chip}
\newacronym{nuca}{NUCA}{non-uniform cache architecture}
\newacronym{numa}{NUMA}{non-uniform memory access}
\newacronym{ooo}{OoO}{out-of-order}
\newacronym{ossu}{OSSU}{Output Spike Speculation Unit}
\newacronym{pc}{PC}{program counter}
\newacronym{pim}{PIM}{Processing-in-Memory}
\newacronym{pe}{PE}{processing element}
\newacronym{pl}{PL}{programmable logic}
\newacronym{pmca}{PMCA}{programmable manycore accelerator}
\newacronym{pulp}{PULP}{Parallel Ultra Low Power}
\newacronym[firstplural=Pretrained Foundation Models]{pfm}{PFM}{Pretrained Foundation Model}
\newacronym[firstplural=Foundation Models]{fm}{FM}{Foundation Model}
\newacronym[firstplural=Vision Transformers]{vit}{ViT}{Vision Transformer}
\newacronym{asr}{ASR}{Automatic Speech Recognition}
\newacronym{raw}{RAW}{read-after-write}
\newacronym{rram}{RRAM}{Resistive Random-Access Memory}
\newacronym{rom}{ROM}{read-only memory}
\newacronym{rmw}{RMW}{read–modify–write}
\newacronym{rob}{ROB}{reorder buffer}
\newacronym{rvwmo}{RVWMO}{RISC-V Weak Memory Ordering}
\newacronym{ro}{RO}{read-only}
\newacronym{rpas}{RPAS}{Random-Projection Attention Speculation}
\newacronym{rtl}{RTL}{register-transfer level}
\newacronym{sbt}{SBT}{static binary translation}
\newacronym{scm}{SCM}{standard cell memory}
\newacronym{sc}{SC}{store-conditional}
\newacronym{sdf}{SDF}{Standard Delay Format}
\newacronym{sdfm}{SDFM}{Sparsity-adaptive Dynamic Frequency Modulation}
\newacronym{simd}{SIMD}{single instruction, multiple data}
\newacronym{simt}{SIMT}{single instruction, multiple thread}
\newacronym{soa}{SoA}{State-of-the-Art}
\newacronym{sm}{SM}{streaming multiprocessor}
\newacronym{soc}{SoC}{system-on-chip}
\newacronym{sprint}{SPRINT}{SParse attention acceleration with appRoximate IN-memory Token pruning}
\newacronym[longplural={scratchpad memories}]{spm}{SPM}{scratchpad memory}
\newacronym{sram}{SRAM}{static random-access memory}
\newacronym{ssa}{SSA}{static single assignment}
\newacronym{ssr}{SSR}{stream semantic register}
\newacronym{stp}{STP}{Sparse Transformer Processor}
\newacronym{tcdm}{TCDM}{tightly-coupled data memory}
\newacronym{tc}{TC}{Tensor Core}
\newacronym{tpc}{TPC}{Tensor Processor Core}
\newacronym{tlp}{TLP}{thread-level parallelism}
\newacronym{tspe}{TSPE}{tensor streaming processing element}
\newacronym{vpu}{VPU}{vector processing unit}
\newacronym{vliw}{VLIW}{very long instruction word}
\newacronym{vnb}{VNB}{von Neumann bottleneck}
\newacronym{war}{WAR}{write-after-read}
\newacronym{waw}{WAW}{write-after-write}
\newacronym{wse}{WSE}{wafer-scale engine}
\newacronym{nar}{NAR}{Non-Autoregressive}
\newacronym{ar}{AR}{Autoregressive}

%% file: Sections/00_Abstract.tex
\begin{abstract}
\glsresetall
Transformer-based foundation models have become crucial for various domains, most notably \gls{nlp} or \gls{cv}. These models are predominantly deployed on high-performance GPUs or hardwired accelerators with highly customized, proprietary instruction sets. Until now, limited attention has been given to RISC-V-based general-purpose platforms. In our work, we present the first end-to-end inference results of transformer models on an open-source many-tiny-core RISC-V platform implementing distributed \softmax{} primitives and leveraging \acrshort{isa} extensions for \acrshort{simd} floating-point operand streaming and instruction repetition, as well as specialized \acrshort{dma} engines to minimize costly main memory accesses and to tolerate their latency. We focus on two foundational transformer topologies, encoder-only and decoder-only models.
For encoder-only models, we demonstrate a speedup of up to 12.8\x between the most optimized implementation and the baseline version. We reach over 79\% FPU utilization and 294 GFLOPS/W, outperforming \gls{soa} accelerators by more than 2\x utilizing the HW platform while achieving comparable throughput per computational unit.
For decoder-only topologies, we achieve 16.1\x speedup in the \gls{nar} mode and up to 35.6\x speedup in the \gls{ar} mode compared to the baseline implementation. Compared to the best SoA dedicated accelerator, we achieve 2.04\x higher FPU utilization.

\end{abstract}

%% file: Sections/01_Introduction.tex
\section{Introduction}
\label{sec:intro}
In the last few years, the field of \gls{ai} has seen a paradigm shift towards transformer-based models such as \glspl{llm} \cite{zhao_survey_2023}, \glspl{vit} \cite{dosovitskiy_image_2020}, and, more in general, \glspl{fm} \cite{zhou_comprehensive_2023}.
Transformer-based \glspl{fm} are large-scale models based on the so-called attention mechanism~\cite{vaswani_attention_2017}, trained on enormous generic datasets in a self-supervised fashion. FMs can then be fine-tuned to perform specialized downstream tasks~\cite{bommasani_opportunities_2022}.
The big breakthrough in \glspl{fm} came with the \gls{gpt}, which exploited the transformer architecture for question-answering problems with natural language.
After the immense success of GPT-3~\cite{brown_language_2020}, transformer-based \glspl{fm} have then been adapted to all aforementioned domains, with models such as \glspl{vit}~\cite{dosovitskiy_image_2020} for computer vision or Whispers \cite{OpenAIWhisper} for \gls{asr}.

The broad adoption of \glspl{fm} makes their efficient execution a high-priority target.
However, the computational patterns at the core of attention layers, i.e., the fundamental building blocks of \glspl{fm}, exhibit several key bottlenecks.
For each input element, these layers compute an \textit{attention score} with respect to every other input obtained through a Softmax-normalized dot-product operation~\cite{vaswani_attention_2017}. Attention complexity scales quadratically with the input sequence length, limiting both the practical size of the models and the length of the inputs they can handle.
Moreover, the internal Softmax is crucial for normalizing the attention scores to a pseudo-probabilistic distribution, but it introduces significant computational challenges. Namely, the computation of exponentials for each attention score prior to normalization can be expensive and susceptible to numerical stability issues, particularly under large-scale or low-precision arithmetic conditions.

At the algorithmic level, these bottlenecks can be addressed with innovations such as sparse and linearized attention mechanisms~\cite{ribar_sparq_2024, katharopoulos_transformers_2020}, as well as approximations and alternatives to the Softmax function~\cite{dao_flashattention-2_2023}. All these approaches aim to mitigate the computational overhead and enhance the model's capacity to handle longer sequences more effectively, thereby extending the practical applications of \glspl{fm} without compromising their performance.

On the HW side, a variety of platforms have been engineered from scratch or enhanced to provide the computational power and memory requirements essential for supporting this new class of models. 
The most commonly used hardware platforms to train and run \glspl{fm} are \glspl{gpu}, such as NVIDIA's A100 and H100~\cite{choquette_nvidia_2021, h100}, coupled with efficient software libraries to limit memory transfers and efficiently distribute the workload.
In a quest to surpass the efficiency of GPUs and primarily address the inference workloads, many new companies have focused on designing specialized \gls{ai} accelerators, such as Cerebras~\cite{dey_cerebras-gpt_2023} and Groq~\cite{abtsSoftwaredefinedTensorStreaming2022}, which employ highly parallel architectures and fully specialized processing elements (typically large arrays of multiply-accumulate units), together with tailored closed-source kernel libraries.
However, accelerators often lack flexibility and cannot compete with GPUs in rapidly adapting to new workloads. 

This work presents the first end-to-end \gls{fm} deployment flow based entirely on an open-source Instruction Set Architecture (ISA), open hardware and open-source software. We target a many-tiny-core general-purpose RISC-V architecture with specialized \riscv{} extensions and advanced \gls{dma} engines, allowing for a highly flexible AI-oriented dataflow. 
This platform adopts a scalable, hierarchical structure that organizes cores into groups of parallel compute clusters. Advanced \gls{dma} engines facilitate high bandwidth heterogeneous memory movements, including core-to-core, cluster-to-cluster, and cluster-to-\gls{hbm} communications, both one-dimensional (1D) and two-dimensional (2D), effectively hiding the latency of costly memory operations~\cite{benz_high-performance_2024}. Each cluster~\cite{zaruba_snitch_2021} comprises eight compute cores with custom \acrshort{isa} extensions for \acrshort{simd}-capable \gls{fp} operand streaming and instruction repetition, as well as one \gls{dma} core. Compute cores are coupled to a 64-bit wide SIMD-capable \gls{fpu} supporting \fpd, \fps, \fph, \texttt{BrainFloat16} (\bfh)~\cite{kalamkar_study_2019}, \fpb (E5M2), and \fpba (E4M3)~\cite{micikevicius_fp8_2022} formats. At the software level, we exploit these hardware capabilities to implement an optimized kernel library for \glspl{fm}, including advanced attention implementations such as FlashAttention-2, fused layers combining head concatenation and input space projection in a logarithmic tree fashion distributed over compute clusters, and low-precision SIMD-based kernels. 

We benchmark our work on five different \glspl{fm}, including decoder-only \glspl{llm}  and encoder-only \gls{vit} models, to demonstrate the flexibility of our HW platform and our SW kernels. Furthermore, we compare against both commercial and academic \gls{soa} architectures. 
We focus on the conventional (quadratic) attention, since algorithmic optimizations (such as linearized attention) are not accuracy-neutral at scale. Furthermore, quadratic attention allows direct comparisons with other state-of-the-art architectures.

The key contributions of this work are the following: 

\begin{itemize}[leftmargin=*]

    \item We provide a comprehensive open-source \gls{fm} library that supports encoder- and decoder-only models, leveraging the HW features and ISA extensions of a scalable RISC-V-based open-source multi-core platform. To improve the performance of our kernels, we leverage advanced \acrshort{dma} engines and demonstrate the benefits of using cluster-to-cluster data transfers, enabling layer fusion and reduction in main memory accesses.
    
    \item We provide a detailed ablation study showing how specialized \riscv \acrshort{isa} extensions for latency-tolerant operand streaming and instruction repetition boost the performance by a factor of up to 35.6\x on \gpt \gls{ar} mode, 16.1\x on the \gpt \gls{nar} mode and 12.8\x for the \gls{vit} model class, respectively, compared to the base \gls{isa}, achieving \acrshort{fpu} utilization above 79\% in \acrshort{nar} mode.
    
    \item We explore the scalability of attention kernels among data precision and the number of available clusters on our HW target. We benchmark our \gls{fm} library at \fpd, \fps, \fph, and \fpb. We also show how different attention block hyperparameters scale compared to the number of cores in terms of throughput (images/s) on all \gls{vit} models. 
    
    \item To the best of our knowledge, we provide the first fully open-source software deployment of \gls{vit} and \gls{llm} models on an open-source \riscv hardware architecture. 
    
    \item Through comprehensive benchmarking, we demonstrate that our end-to-end inference engine outperforms \gls{soa} platforms regarding HW utilization while maintaining the full flexibility of a many-core RISC-V architecture. We outperform \gls{soa} accelerators by up to 8\x in terms of \gls{fpu} utilization, with a minimum speedup of 1.81\x compared to the best competitor. 
\end{itemize}

The remainder of the manuscript is organized as follows. Sections ~\ref{sec:background} and ~\ref{sec:related} provide background and review of related work in full-stack \gls{fm} deployment flows. Section~\ref{sec:hw} offers an in-depth description of our target hardware platform and Section~\ref{sec:library} details our software library. Section~\ref{sec:results} presents comprehensive findings and the results of our benchmarking experiments. Finally, Section~\ref{sec:conclusion} summarizes our study with concluding remarks.

%% file: Sections/02_Background.tex
\section{Background}
\label{sec:background}
\subsection{Attention Kernel}
\label{subsec:attention}
\begin{figure}[t]
    \centering
    \includegraphics[width=0.95\linewidth]{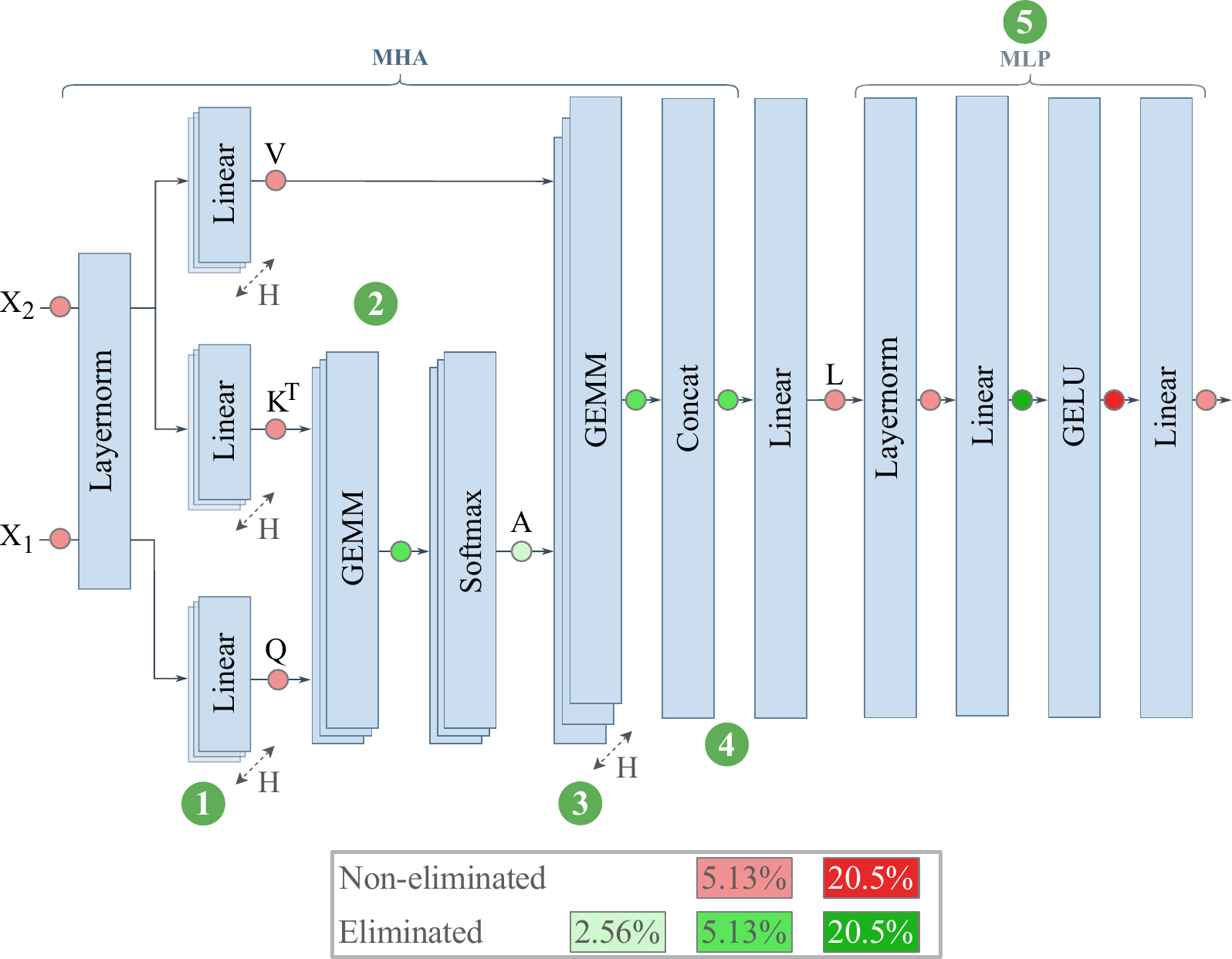}
    \caption{Block topology of the basic Attention layer. The arrows are annotated with the percentage of the memory transfers needed for the specific tensor over the total number of transfers needed by the block, considering the \gpt-j model in \gls{nar} mode and a sequence length of 2048. Red dots represent data reads from HBM of our implementation. Green ones are transfers done only at the cluster level.}
    \label{fig:attention}
\end{figure} 
Transformers-based \glspl{fm} are constructed by stacking multiple \textit{attention} blocks, which correlate each element of a sequence $\mathbf{X_{1}}$ (e.g., language tokens or image patches) with all elements of another sequence $\mathbf{X_{2}}$~\cite{vaswani_attention_2017,devlin_bert_2019}.
Notably, in many cases, the two sequences coincide ($\mathbf{X_{1}} = \mathbf{X_{2}}$), and the operation is denoted as \textit{self-attention}, as opposed to \textit{cross-attention} when they differ.
The ability of attention layers to dynamically \emph{focus} on different input data segments has been shown to improve performance across diverse tasks, thus becoming a fundamental component in the architecture of modern \gls{soa} neural networks.

A scheme depicting a basic attention block is shown in Figure~\ref{fig:attention}. The vectors that compose the input sequences organized as rows of the $\mathbf{X_{1}}$ and $\mathbf{X_{2}}$ tensors are first projected through learned linear transformations \trafocircle{1} to form three new matrices: the \gls{q}, obtained from $\mathbf{X_{1}}$, and the \gls{k} and \gls{v}, derived from $\mathbf{X_{2}}$. Mathematically:

\begin{equation}
\label{eq:proj}
\mathbf{Q} = \mathbf{X_{1}}\mathbf{W}_{\text{Q}},\ \mathbf{K} = \mathbf{X_{2}}\mathbf{W}_{\text{K}},\ \mathbf{V} = \mathbf{X_{2}}\mathbf{W}_{\text{V}}
\end{equation}
Where $\mathbf{W}_{\text{Q, K, V}}$ have dimensions E $\times$ P, which represent the \gls{e} space (i.e., the number of input features) and the \gls{p} space, respectively~\cite{vaswani_attention_2017}.

The attention weights (or scores) $\mathbf{A}$ are then calculated as the scaled dot product of matrix \gls{q} and \gls{k}, normalized using a \softmax{} function to convert them into pseudo-probabilities \trafocircle{2}. The scaling factor $\sqrt{\mathbf{P}}$, where $\mathbf{P}$ corresponds to the dimension of the key vector, ensures that the dot products do not become excessively large, which aids in maintaining numerical stability during training. The normalized scores represent the degree to which each element of $\mathbf{X_{1}}$ should attend to all elements of $\mathbf{X_{2}}$, directing the network's \emph{focus} towards more relevant parts of the input.
Accordingly, a new sequence is produced in which each element is a weighted sum of the \gls{v} vectors, where the weights are the \softmax-normalized scores. Formally:
\begin{equation}
\label{eq:attn}
\text{Attention}(\mathbf{Q}, \mathbf{K}, \mathbf{V})
\doteq
\mathbf{A} \mathbf{V}
\doteq
\softmax_\text{over keys} \left(\frac{\mathbf{Q}\mathbf{K}^\text{T}}{\sqrt{\mathbf{P}}} \right)\mathbf{V}
\end{equation} 

The commonly used \gls{mha} extends this basic layer to enhance the model's capability to process information from multiple representation subspaces simultaneously~\cite{vaswani_attention_2017}. It allows the model to capture various aspects of the input data in parallel \textit{heads}, each performing the attention process independently with different learned linear transformations. After each head computes its attention outputs independently, the results are concatenated and once again linearly transformed \trafocircle{4}. This step combines the different representations of each head, forming a unified output. 
The output vectors of the \gls{mha} block are normalized with a Layernorm operation and then processed by a \gls{mlp} \trafocircle{5}. The first dense layer of the \gls{mlp} expands the input feature dimension to the hidden dimension. The subsequent \gls{gelu} activation function introduces non-linearity, which is essential for learning more complex patterns.  The final linear layer projects the expanded representations back to the original input space \gls{e}. 

While new attention block variants are arising, such as the \gls{gqa}~\cite{touvron_llama_2023}, we describe and employ classical attention for our experiments to enable fair comparisons on the same workload with SoA hardware. However, the findings of our work can be applied to any other Transformer-based architecture.  

\begin{figure}[t]
    \centering
    \includegraphics[width=0.99\linewidth]{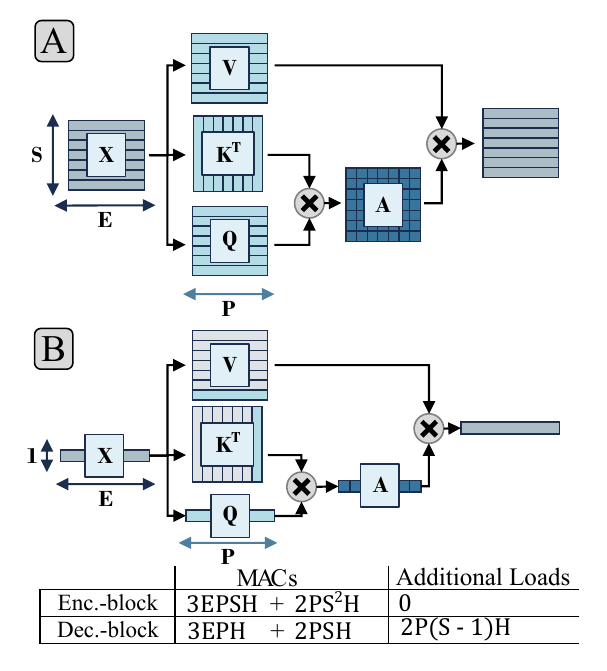}
    \caption{Operation performed by the fundamental ViT and GPT blocks.}
    \label{fig:enc_dec}
\end{figure}

\subsection{Foundation Models}
\label{subsec:fm}
Transformers revolutionized \gls{nlp}~\cite{vaswani_attention_2017} and later, \gls{cv}~\cite{dosovitskiy_image_2020}, text generation~\cite{dettmers_gpt3int8_2022}, and other domains, thanks to their excellent representation capability. As a consequence of this remarkable performance, coupled with the difficulty of training large transformers from scratch for each new task and the effectiveness of newly developed label-free training schemes, \glspl{fm}~\cite{bommasani_opportunities_2022} have emerged as the most natural paradigm for training and deploying transformers, leveraging vast amounts of data~\cite{dettmers_gpt3int8_2022, dosovitskiy_image_2020}. 
\glspl{fm} undergo extensive pre-training on large and diverse datasets, often through self-supervised procedures, which helps them develop a generalized and wide-range understanding of language patterns, visual information, audio, and many more mixed modalities, depending on the nature of training data. Once pre-trained, these models can be fine-tuned at a much lower cost (in terms of data and computing) to perform specific downstream tasks accurately.

The large majority of \gls{soa} \glspl{fm} belong to one of two big classes. \textit{Encoder-only} models, which include classifiers such as \glspl{vit}, are feed-forward neural networks that simultaneously receive the entire input sequence and generate their output in a single inference pass. \textit{Decoder-only} models, in contrast, are mainly used for generative tasks, typically in an autoregressive fashion. In this case, the final output is generated through multiple iterations of the network, each fed with the output of the previous one.
In our work, we describe a general \gls{fm} kernel library and a flexible hardware platform able to optimize both model families. We use the dominant \gls{vit} and \gls{gpt} as representatives of the two categories to demonstrate our results.
\begin{description}
    \item[\gls{vit}-based \glspl{fm}] The \gls{vit} encoder-only architecture exploits a first convolutional layer with a stride equal to the filter dimension to create a linearized series of patches used as the input sequence. The heart of the network constitutes \gls{mhsa} blocks that correlate all the input tokens with themselves, therefore having $\mathbf{X_1} = \mathbf{X_2}$. Fig. \ref{fig:enc_dec}-A depicts the matrix operations executed and the tensor dimensions for the key blocks of the \gls{mhsa} section. The operations are quadratically proportional to the input sequence length $S$ and linearly proportional to $P$ and $E$. All involved operations are on multi-dimensional matrices, and a single pass through all the network blocks produces the desired classification.
    As reference models for this family, we considered \gls{vit}-\{Base, Large, Huge\}, denoted with suffixes -B, -L, and -H in the following, given their maturity and widespread adoption~\cite{dosovitskiy_image_2020}. Although more recent models have been developed, the basic building blocks remained similar, and, therefore, our library can be identically employed for their deployment~\cite{wu_visual_2020,wu_cvt_2021}. 
    \item[\gls{gpt}-based \glspl{fm}] Decoder-only models of the \gls{gpt} family are employed as \glspl{llm}. They elaborate sequences of language \textit{tokens} (words or portions thereof) from a dictionary. In their most basic use case, given an initial prompt, they are tasked to generate a highly likely continuation. Computationally, there are two crucial differences compared to the encoder model: First, the attention is causal, which means that the new tokens can only attend to the previous ones in the sequence and not to subsequent ones. Second, a single new token and not the entire output is generated at each pass through the network (see Fig. \ref{fig:enc_dec}-B). For instance, to generate the reply to the question \emph{What is your name?}, the model is first provided with the tokenized question as a prompt. In the first iteration, the initial token of the continuation (e.g., \emph{I}) is generated. Then, such a token is appended to the input, and the network is invoked again, producing the following output token (e.g., \emph{am}). This procedure repeats until a special end-of-sentence token is produced or another stopping criterion is met (e.g., a maximum length).
    Compared to encoder-only models,  we observe a shift from matrix-matrix operations towards matrix-vector operations, which are less compute-intensive. In fact, except for the initial network invocation (prompt processing or pre-fill step), a single new query vector per layer is processed at each subsequent step, computing its attention weights against all previous keys and values. Note that the K and V projections involve just a vector-matrix multiplication since the keys and values relative to tokens processed in previous inference passes can be stored in memory to avoid re-computation in the so-called \textit{KV cache} \cite{pope_efficiently_2023}. These differences will be analyzed in detail in Sec. \ref{sec:results}.
\end{description}

\subsection{Transformer Optimization}
\label{sec:opt_background}
In the domain of \glspl{fm}, optimizing for computational efficiency without compromising performance is a crucial challenge, given the high amount of resources required by these models and the recent trends towards embedding them on constrained mobile or edge devices~\cite{liu_mobilellm_2024}.

An important category of optimizations in the literature looks at the network topology, either changing the attention block's structure or simplifying the overall network. For instance, a series of works \cite{ma_luna_2021, wang_linformer_2020} proposed simplifying the attention computation by either low-rank approximation or fixing a dimension, reducing the complexity with respect to $S$ from quadratic to linear. Other approaches focus on sharing the $K$ and $V$ matrices across multiple heads, obtaining the so-called \gls{gqa} found in more recent LLaMa models \cite{touvron_llama_2023}, or applying sparsity to the projection weights to reduce the overall memory occupation.
Some works specifically focus on \glspl{llm} by reducing the KV-cache size, either by compression or by evicting some unimportant tokens \cite{liu_scissorhands_2023}.
All these optimizations are orthogonal to our work, and indeed, the flexibility of our architecture enables adopting all these algorithmic optimizations once they are fully proven to be robust and competitive in accuracy.

Consequently, we focus our analysis on optimizations concerning canonical attention, the most widespread ViT and GPT transformer architecture, and the corresponding data precision.
Despite many studies exploring the possibility of using low-precision \textit{integer} formats for \glspl{fm}, such as 8-bit \cite{dettmers_gpt3int8_2022} or 4-bit \cite{dettmers_qlora_2023}, these quantization schemes inevitably lead to a non-negligible loss in accuracy. In contrast, the research on \fpb quantization, as documented in \cite{micikevicius_fp8_2022} by NVIDIA researchers' testing across various GPT, encoder-decoder, and BERT models, shows that low bit width floating point formats can often maintain identical performance without any loss, while significantly enhancing computational efficiency.
Hence, in this work, our efforts have been concentrated on testing the performance scalability of our hardware and software stack considering precisions ranging from 64-bit down to 8-bit floating-point.
A second key optimization knob to improve \gls{fm} performance consists in reducing the number of costly accesses to the main memory. To this end, we exploit the \emph{FlashAttention-2} implementation~\cite{dao_flashattention-2_2023} and a fused linear concatenation layer that leverages binary cluster-to-cluster sum reduction to concatenate the attention heads and project them to the original input space dimension. Both techniques minimize frequent communication between the main \gls{hbm} and the on-chip \gls{spm} during the processing of attention blocks and can be applied to any Transformer-based \gls{fm}. While both have already been explored in literature, we adapt them to our HW platform, exploiting its unique features.

%% file: Sections/03_RelatedWork.tex
\section{Related Work}
\label{sec:related}
\input{tables/01_Table_SoA}
\looseness=-1000
While \glspl{gpu} are still the most commonly used hardware platforms for \gls{ai} workloads~\cite{choquette_nvidia_2021, noauthor_nvidia_nodate,smith_amd_2022}, various companies have designed their specialized accelerators in the last few years~\cite{lie_cerebras_2023,systems_sambanova_nodate,knowles_graphcore_2021, noauthor_intel_nodate}.
Table~\ref{tab:soa} presents an overview of some of the most recent platforms specifically designed for \gls{ai} training and inference, often focusing on \glspl{llm} and \glspl{fm}.

A common denominator of those architectures is the replication of a base compute cluster combining multiple \glspl{pe} with an L1 cache or \gls{spm} via a low-latency interconnect.
The replicated clusters are interconnected by a latency-tolerant global \gls{noc} and rely either entirely on external main memory or on some distributed on-chip cluster memories.

For example, \nvidia{}’s A100~\cite{choquette_nvidia_2021} groups multiple Texture Processing Clusters with 108 \glspl{sm} into GPU Processing Clusters.
Each \gls{sm} combines four warps with an L1-data cache and a shared memory of \SI{164}{\kilo\byte}, whereas each warp combines a tensor core, 16 \texttt{INT32} and \texttt{FP32} cores, and eight \texttt{FP64} cores with a shared \SI{64}{\kilo\byte} register file.
The tensor core supports \texttt{FP16}, \texttt{BF16}, \texttt{FP32}, \texttt{FP64}, \texttt{INT8}, \texttt{INT4}, and binary formats, and its 32-thread granularity enables 256 \texttt{FP16}/\texttt{FP32} \gls{fma} operations per cycle, corresponding to the computation of an 8\x{}4\x{}8 mixed-precision matrix multiplication.

The H100~\cite{noauthor_nvidia_nodate} GPU doubles the HBM memory of the A100 and increases the number of total cores from 6912 to 16896, and introduces support for \fpb. Further, it includes a Transformer Engine (TE) optimized for handling multi-trillion parameter models, significantly enhancing performance for \glspl{fm}. The engine supports Automatic Mixed Precision (AMP), which dynamically adjusts the precision of computations during training and inference. Additionally, the TE supports fused operations, combining multiple computational steps into a single kernel. The architecture of the TE is designed to scale efficiently across multiple \glspl{gpu}, allowing for parallel training and inference of large models, further reducing training times. This scalability is complemented by an advanced software stack optimized for transformer workloads, including libraries like \texttt{NVIDIA CUDA}, \texttt{cuBLAS}, and \texttt{cuDNN}.

SambaNova DataScale SN30~\cite{systems_sambanova_nodate}  uses second-generation Reconfigurable Dataflow Units (RDU) as \glspl{pe}. It uses a \SI{7}{\nano\meter} process node and supports \bfh and \fps data formats.
Cerebras has scaled its accelerator architecture to a wafer-sized chip called \gls{wse}~\cite{lie_cerebras_2023}.
The second generation, \gls{wse}-2, with \SI{46000}{\milli\metre\squared}, is the largest processor ever built and combines 850k \gls{ai} cores with a wafer-scale high-bandwidth, low-latency fabric arranged in a 2D mesh.
Each general-purpose core combines four 16-bit \gls{fmac} units, 8\x \SI{6}{\kilo\byte} \gls{sram} banks (total \SI{48}{\kilo\byte}), and \SI{256}{\byte} of software-managed cache for often accessed data structures such as accumulators.

GROQ’s \gls{lpu} architecture~\cite{abtsSoftwaredefinedTensorStreaming2022}, taped out in \SI{14}{\nano\meter}, employs a linear array of \glspl{tspe} interleaved with memory units. It is designed for deterministic performance and elimitates control flow variability to maximize data throughput. The LPU supports \fps, \fph, and \texttt{INT8}. Each LPU comprises \SI{220}{\mega\byte} of on-chip \acrshort{sram}, divided across superlanes, enabling high performance without the need for external \acrshort{dram}. The LPU can compute more than 400000 integer \gls{mac} operations per cycle, achieving up to 1000 TOPS. By having the compiler explicitly control all aspects of program execution, including instruction scheduling, data movement, and resource management, runtime variability is eliminated.

The Habana Gaudi2~\cite{noauthor_intel_nodate} processor features
two Matrix Multiplication Engines (MME), 24 \glspl{tpc}, and \SI{96}{\giga\byte} of \gls{hbm}, providing a robust platform for efficient scale training. It supports \fpb, \fph, and \fps data formats.
Lastly, the AMD MI250 \cite{smith_amd_2022} utilizes a \SI{6}{\nano\meter} process with \SI{128}{\giga\byte} of HBM2e, focused on high bandwidth and large buffer capacities essential for handling large datasets efficiently. It supports \fph, \fps, and \fpd, together with low-precision integer formats, aligning with the needs of diverse AI training and inference workloads.

In contrast to large industrial products, academic research has focused on smaller-scale hardware and models, investigating specific optimizations. 
For instance, AccelTran~\cite{tuliAccelTranSparsityAwareAccelerator2023} focuses on sparse transformers, using a dynamic inference scheme to prune activations at runtime.
In most academic works~\cite{tuliAccelTranSparsityAwareAccelerator2023, wang_energy-efficient_2023, kim20CTransformer6182024, shanmugasundaramFreFlexHighPerformanceProcessor2024, qinAyakaVersatileTransformer2024}, the authors primarily focus on specific hardware optimizations to support different integer quantization schemes that can limit the applicability to high-accurate LLMs.
Additionally, they propose architectures tailored to specific models, such as fusions between Transformers and DNNs or Spiking Neural Networks (SNNs)~\cite{kim20CTransformer6182024}.
Tambe \etal{}~\cite{tambe2212nm182023} are the only ones employing floating point data types. However, the authors combine floating-point with a series of other software optimizations that could limit the utilization of their work for huge LLMs. These optimizations include token entropy-based early exit, \gls{mp} (\texttt{FP4}/\texttt{FP8}) prediction, and sparsity.
Until now, no open-source general-purpose academic hardware platforms have demonstrated end-to-end execution of widespread foundation models like GPT and ViT. 

To fill this gap, in this work, we explore an \textit{open-source}, scalable architecture comprising a hierarchical grouping of four \riscv{} multi-core Snitch compute clusters~\cite{occamy_vlsi_2024}, achieving an inter-cluster bandwidth of 64B/cycle while sharing 64B/cycle bandwidth to another hierarchy stage of cluster grouping. We exploit on-chip interconnect to enable flexible cluster-to-cluster direct communication without passing through main memory. This capability is not fully supported even in the most advanced commercial platforms, where direct on-chip communication is either restricted to a very rigid dataflow (e.g., Groq~\cite{abtsSoftwaredefinedTensorStreaming2022}) or limited to an expensive shuffle (e.g., in NVIDIA H100~\cite{h100}).

%% file: tables/01_Table_SoA.tex
\begin{table*}[t]
\caption{State-of-the-art platforms for \glspl{fm} execution.}
\label{tab:soa}
\centering
\renewcommand{\arraystretch}{1.15}
\resizebox{\linewidth}{!}{%
\begin{tabular}{lllllll}
\hline
\multicolumn{1}{l|}{Platform}                                                            & Technlogy                                                  & \begin{tabular}[c]{@{}l@{}}Max.\\      Frequency\end{tabular} & Memory                                                                                        & Cores                                                                        & \begin{tabular}[c]{@{}l@{}}Data\\      Format\end{tabular}                                                         & \begin{tabular}[c]{@{}l@{}}Special\\      Features\end{tabular}                \\ \hline\hline
\multicolumn{7}{l}{Commercial Platforms}                                                                                                                                                                                                                                                                                                                                                                                                                                                                                                                                                                   \\ \hline\hline
\multicolumn{1}{l|}{A100~\cite{choquette_nvidia_2021}}                                                                & \begin{tabular}[c]{@{}l@{}}\SI{7}{\nano\meter} N7 \\ TSMC \end{tabular}                                                     & \SI{1.41}{\giga\hertz}                                        & \begin{tabular}[c]{@{}l@{}}\SI{164}{\kilo\byte} SPM\\      + \SI{40}{\giga\byte} \gls{hbm}\end{tabular}                    & \begin{tabular}[c]{@{}l@{}}6912 FP32\\      + 432 \acrshort{tc}s\end{tabular}  & \begin{tabular}[c]{@{}l@{}}\texttt{FP8}/\texttt{FP16}/\texttt{FP32}/\\ \texttt{FP64}/\texttt{INT8}/\texttt{INT4}\end{tabular}    & \acrshort{tc}s  \\ \hline 
\multicolumn{1}{l|}{H100~\cite{noauthor_nvidia_nodate}}                                                                & \begin{tabular}[c]{@{}l@{}}\SI{4}{\nano\meter} N4 \\ TSMC \end{tabular}                                                     & \SI{1.78}{\giga\hertz}                                        & \begin{tabular}[c]{@{}l@{}}\SI{256}{\kilo\byte} SPM\\      + \SI{80}{\giga\byte} \gls{hbm}\end{tabular}              & \begin{tabular}[c]{@{}l@{}}16896 FP32\\      + 528 \acrshort{tc}s\end{tabular} & \begin{tabular}[c]{@{}l@{}}\texttt{FP8}/\texttt{FP16}/\texttt{FP32}/\\ \texttt{FP64}/\texttt{INT8}/\texttt{INT4}\end{tabular}    &  \begin{tabular}[c]{@{}l@{}}Transformer Engine, \\ Tensor Cores\end{tabular}                                                                    \\ \hline 
\multicolumn{1}{l|}{\begin{tabular}[c]{@{}l@{}}SambaNova \\ DataScale SN30~\cite{systems_sambanova_nodate}\end{tabular}} & \begin{tabular}[c]{@{}l@{}}\SI{7}{\nano\meter} N7 \\ TSMC \end{tabular}                                                     & -                                                             & \begin{tabular}[c]{@{}l@{}}\SI{640}{\mega\byte} \acrshort{sram}\\      + \SI{1}{\tera\byte} \acrshort{dram}\end{tabular}   & 1280 PCUs                                                                    & \begin{tabular}[c]{@{}l@{}}\texttt{BF16}/\texttt{FP32}/\texttt{INT32}/\\ \texttt{INT16}/\texttt{INT8}\end{tabular} & Units (PCUs)                                                                   \\ \hline
\multicolumn{1}{l|}{Cerebras CS-2~\cite{lie_cerebras_2023}}                                                       & \begin{tabular}[c]{@{}l@{}}\SI{7}{\nano\meter} N7 \\ TSMC \end{tabular}                                                    & \SI{1.1}{\giga\hertz}                                         & \begin{tabular}[c]{@{}l@{}}\SI{40}{\giga\byte} \acrshort{sram}\\      + \SI{1}{\tera\byte} \acrshort{dram}\end{tabular}    & 850'000                                                                      & \texttt{FP16}/\texttt{FP32}                                                                                        & \acrshort{wse}                                                             \\ \hline
\multicolumn{1}{l|}{GROQ~\cite{abtsSoftwaredefinedTensorStreaming2022}}     &   \begin{tabular}[c]{@{}l@{}}\SI{14}{\nano\meter} \\ GF\end{tabular}   & \SI{900}{\mega\hertz}  & \begin{tabular}[c]{@{}l@{}}\SI{2.3}{\tera\byte} \acrshort{sram}\end{tabular} &   10440 \acrshort{tspe}s  &      \begin{tabular}[c]{@{}l@{}}\texttt{FP16}/\texttt{FP32}/\texttt{INT8}\end{tabular}              &  \begin{tabular}[c]{@{}l@{}}\acrshort{lpu}\end{tabular}  \\ \hline
\multicolumn{1}{l|}{Habana Gaudi2~\cite{noauthor_intel_nodate}}                                                       & \begin{tabular}[c]{@{}l@{}}\SI{7}{\nano\meter} \\ Intel \end{tabular}                                                 & -                                                             & \begin{tabular}[c]{@{}l@{}}\SI{48}{\mega\byte} \acrshort{sram}\\      + \SI{96}{\giga\byte} \acrshort{hbm}\end{tabular}   & \begin{tabular}[c]{@{}l@{}}24 TPC\\      + 2 MME\end{tabular}                & \texttt{FP8}/\texttt{FP16}/\texttt{FP32}                                                                           & Tensor Processor Core                                                          \\ \hline
\multicolumn{1}{l|}{AMD MI250~\cite{smith_amd_2022}}                                                            & \begin{tabular}[c]{@{}l@{}}\SI{6}{\nano\meter} FinFET \\ TSMC\end{tabular} & \SI{1.6}{\giga\hertz}                                         & \begin{tabular}[c]{@{}l@{}}\SI{8}{\mega\byte} \acrshort{sram}\\      + \SI{128}{\giga\byte} \gls{hbm}\end{tabular}        &        \begin{tabular}[c]{@{}l@{}}4096 Stream\\      Processors\end{tabular}                                                                      & \begin{tabular}[c]{@{}l@{}}\texttt{FP16}/\texttt{FP32}/\texttt{FP64}/\\ \texttt{INT8}/\texttt{INT4}\end{tabular}   & AMD CDNA 2                                                                     \\ \hline\hline
\multicolumn{7}{l}{Academic Platforms}                                                                                                                 \\ \hline\hline

\multicolumn{1}{l|}{AccelTran~\cite{tuliAccelTranSparsityAwareAccelerator2023}}    &   \begin{tabular}[c]{@{}l@{}}\SI{14}{\nano\meter} FinFet \\ Intel \end{tabular} &  \SI{700}{\mega\hertz}    &\begin{tabular}[c]{@{}l@{}} \SI{13}{\mega\byte} 3D RRAM \\ + \SI{16}{\giga\byte} LP-DDR3\end{tabular}  &  \begin{tabular}[c]{@{}l@{}}64 \acrshort{pe}s \end{tabular}   & 20-bit fixed-point &   \begin{tabular}[c]{@{}l@{}} DynaTran \\ + dynamic data flow\end{tabular}  \\ \hline




\multicolumn{1}{l|}{Wang \etal{} \cite{wang_energy-efficient_2023}}                                               &         \SI{28}{\nano\meter} CMOS                                                  &                                 \SI{510}{\mega\hertz}                             &    \SI{336}{\kilo\byte} \acrshort{sram} &  \begin{tabular}[c]{@{}l@{}} 4 Cores \\ (32 \acrshort{pe} Lines)\end{tabular}                   &   \texttt{INT12}   &        Approx. Attention                                                                        \\ \hline
\multicolumn{1}{l|}{Kim \etal{} \cite{kim20CTransformer6182024}}    &   \begin{tabular}[c]{@{}l@{}}\SI{28}{\nano\meter} CMOS \\ TSMC \end{tabular}  & \SI{200}{\mega\hertz}    &   \SI{500}{\kilo\byte} SPM  &  \begin{tabular}[c]{@{}l@{}}\acrshort{hdsc} + \acrshort{hmau} \\ \acrshort{ossu}, \acrshort{iwgu}\end{tabular}   & \texttt{INT8} &   \begin{tabular}[c]{@{}l@{}} SNN/DNN transformer \\ Big-little network\end{tabular}  \\ \hline

\multicolumn{1}{l|}{FreFlex~\cite{shanmugasundaramFreFlexHighPerformanceProcessor2024}}    &   \begin{tabular}[c]{@{}l@{}}\SI{28}{\nano\meter} CMOS \\ TSMC \end{tabular}  &  \SI{1.1}{\giga\hertz}    &\begin{tabular}[c]{@{}l@{}} \SI{64}{\kilo\byte} SRAM\end{tabular}  &  \begin{tabular}[c]{@{}l@{}}32 x 16 PE array \end{tabular}   & \texttt{INT8} &   \begin{tabular}[c]{@{}l@{}}  \acrshort{sdfm}\end{tabular}  \\ \hline

\multicolumn{1}{l|}{Ayaka~\cite{qinAyakaVersatileTransformer2024}}    &   \begin{tabular}[c]{@{}l@{}}\SI{28}{\nano\meter} CMOS \\ TSMC \end{tabular}  &  \SI{430}{\mega\hertz}    &\begin{tabular}[c]{@{}l@{}} \SI{544}{\kilo\byte} SRAM\end{tabular}  &  \begin{tabular}[c]{@{}l@{}}\acrshort{rpas} unit \\ + transformer core \end{tabular}   & \texttt{INT8}/\texttt{INT16} &   \begin{tabular}[c]{@{}l@{}} \acrshort{csp} + \acrshort{hdpe}\end{tabular}  \\ \hline

\multicolumn{1}{l|}{Tambe \etal{}~\cite{tambe2212nm182023}}    &   \begin{tabular}[c]{@{}l@{}}\SI{12}{\nano\meter} FinFET \\ N/A \end{tabular}  &  \SI{717}{\mega\hertz}   &\begin{tabular}[c]{@{}l@{}} \SI{647}{\kilo\byte} SRAM\end{tabular}  &  \begin{tabular}[c]{@{}l@{}} \acrshort{stp} \end{tabular}   & \texttt{FP4}/\texttt{FP8} &   \begin{tabular}[c]{@{}l@{}} \acrshort{mp} \acrshort{mac} unit\end{tabular}  \\ \hline

\multicolumn{1}{l|}{Ours}                                                                & \begin{tabular}[c]{@{}l@{}}\SI{12}{\nano\meter} FDX \\ GF \end{tabular}                                                   & \SI{1}{\giga\hertz}                                           & \begin{tabular}[c]{@{}l@{}}\SI{128}{\kilo\byte} SPM \\      + \SI{16}{\giga\byte} \gls{hbm}\end{tabular}            & \begin{tabular}[c]{@{}l@{}}16 clusters\\      9 cores/cluster\end{tabular}   & \begin{tabular}[c]{@{}l@{}}\texttt{FP8}/\texttt{FP16}/\\ \texttt{FP32}/\texttt{FP64}\end{tabular}                  & \begin{tabular}[c]{@{}l@{}}MiniFloat \gls{isa}, \\ cluster2cluster communication \\ \glspl{ssr}\end{tabular} \\ \hline
\end{tabular}}
\end{table*}

%% file: Sections/04_HW_platform.tex
\section{Many-Core RISC-V Hardware Platform}
\label{sec:hw}
In this work, we optimize \gls{fm} inference on an open-source, scalable, many-core architecture optimized for \gls{fp}-intensive workloads with \gls{ml}-centric \gls{isa} extensions. The foundation of the hardware platform is the \textit{Snitch} cluster \cite{zaruba_snitch_2021} described in \cref{subsec:snitch-cluster} and shown in \autoref{fig:architecture}, which is replicated and implemented in a hierarchical multi-cluster configuration, as discussed in \cref{subsec:multi-cluster-archi} and shown in \autoref{fig:memory}.

\subsection{Compute Cluster Architecture}
\label{subsec:snitch-cluster}

\begin{figure}
  \centering
  \includegraphics[width=\linewidth]{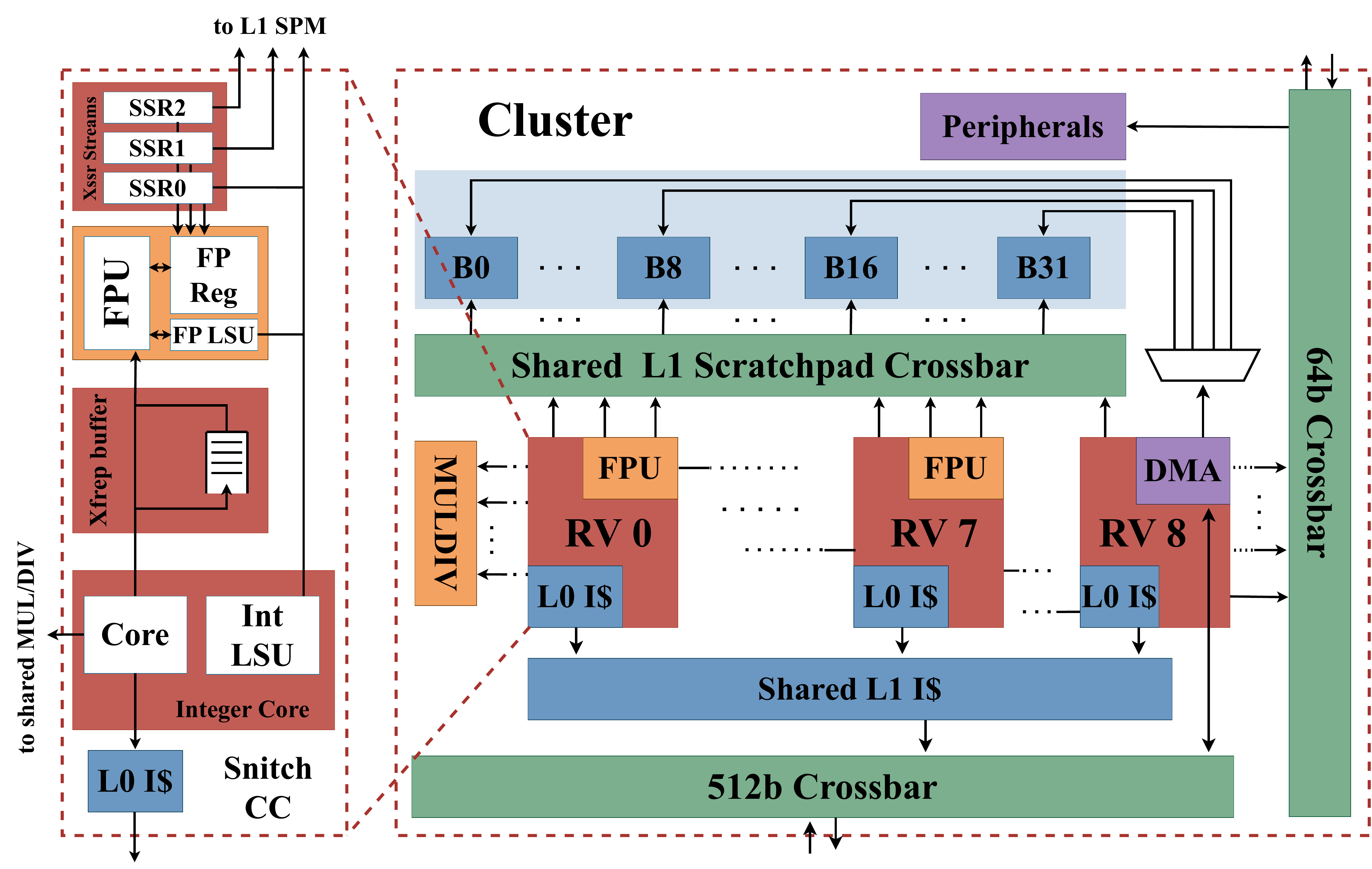}
  \caption{Architecture of the \riscv{} compute cluster with ISA extension \texttt{Xfrep} and \texttt{Xssr}.}
  \label{fig:architecture}
\end{figure}
The compute cluster is the basic block of our scalable HW architecture~\cite{zaruba_snitch_2021}.
\autoref{fig:architecture} illustrates the architecture of the cluster, which is composed of 9 tiny 32-bit in-order integer \riscv{} cores called Snitch~\cite{zaruba_snitch_2021}, each coupled with an L0 instruction cache and a 64-bit wide \gls{simd}-capable \gls{fpu}, supporting a wide range of \gls{fp} formats and mixed-precision operations~\cite{bertaccini_minifloats_2024}.
Additionally, the cores share an integer division/square-root unit and a two-way \SI{8}{\kilo\byte} L1 \gls{idol}.
To achieve \acrshort{fpu} utilization rates above 50\% with Snitch's lean single-issue pipeline, the architecture integrates two extensions~\cite{zaruba_snitch_2021}: \texttt{Xssr} and \texttt{Xfrep}.
The \texttt{Xssr} augments three \gls{fp} register file entries to \glspl{ssr} with capabilities for address generation, enabling latency-tolerant data prefetching.
The \texttt{Xfrep} implements an instruction repetition buffer called \gls{frep}, reducing the pressure on the L1 instruction cache and preventing the overhead of loop handling. 
The joint effect of these extensions is that the inner loop, e.g., a dot product operator, only contains the FMA instruction. Hence, FPU utilization can rise to the 90\%-region.

Eight of these cores are in charge of parallel workloads, while the ninth core is used for cluster coordination and controls a \gls{dma} unit with hardware support for 1D and 2D data transfers.
All cores share a tightly coupled 128kB L1 \gls{spm} via a single-cycle interconnect.
The \gls{spm} itself is organized as 32\x{} 64-bit interleaved memory banks.
The \gls{dma} unit can access eight banks in parallel via a dedicated 512-bit interconnect and shares a duplex 512-bit \gls{axi} crossbar to the higher-level memory system with the \gls{idol}.
The cores can access the global memory system over a duplex 64-bit AXI crossbar.
Furthermore, all cores support atomic operations and use hardware barriers for synchronization.
In Sec. \ref{sec:library}, we will describe how the peculiar HW features are exploited in our kernel library for \glspl{fm}.

\subsubsection{Low and Mixed-Precision Floating-Point Support}
Low-precision data types allow for reduced memory footprints and more energy-efficient inferences for ML models. For these reasons, \snitch's \glspl{fpu} supports a wide set of \gls{fp} formats: \fpd, \fps, two 16-bit \gls{fp} data types (\fph, \bfh), and two 8-bit \gls{fp} data types (\fpb, \fpba). The \fpb and \fpba formats include a 2-bit/3-bit mantissa field and a 5-bit/4-bit exponent field, respectively.
By leveraging the wide set of \gls{fp} formats, the computation precision can be tuned to the specific application requirements, achieving higher performance and energy efficiency. The Snitch cluster has a theoretical peak performance of 16, 32, 64, and 128 FLOP/cycle, respectively for \fpd, \fps, \fph/\bfh, and \fpb/\fpba, when parallelizing the computation over 8 cores (where \hbox{1 fused multiply-add = 2 FLOP}).
Furthermore, Snitch's \gls{fpu} supports \gls{simd} short widening dot-product instructions~\cite{bertaccini_minifloats_2024} working on 8 or 16-bit inputs and accumulating with 16 or 32 bits, respectively. These mixed-precision instructions compute the equivalent of two cascaded widening fused multiply-add. They ensure the speedup enabled by the lower-precision inputs even when accumulating at higher precision while retaining a higher accuracy in long accumulations (for example, in GEMM kernels). 
In our work, we explore the utilization of different data precisions, showing the speedup achievable with a dedicated \gls{fm} library capable of exploiting all the \gls{simd} operations and maintaining higher precision where needed (e.g., softmax).

\subsection{Scalable Multi-Cluster Architecture}
\label{subsec:multi-cluster-archi}
\begin{figure}
    \centering
    \includegraphics[width=\linewidth]{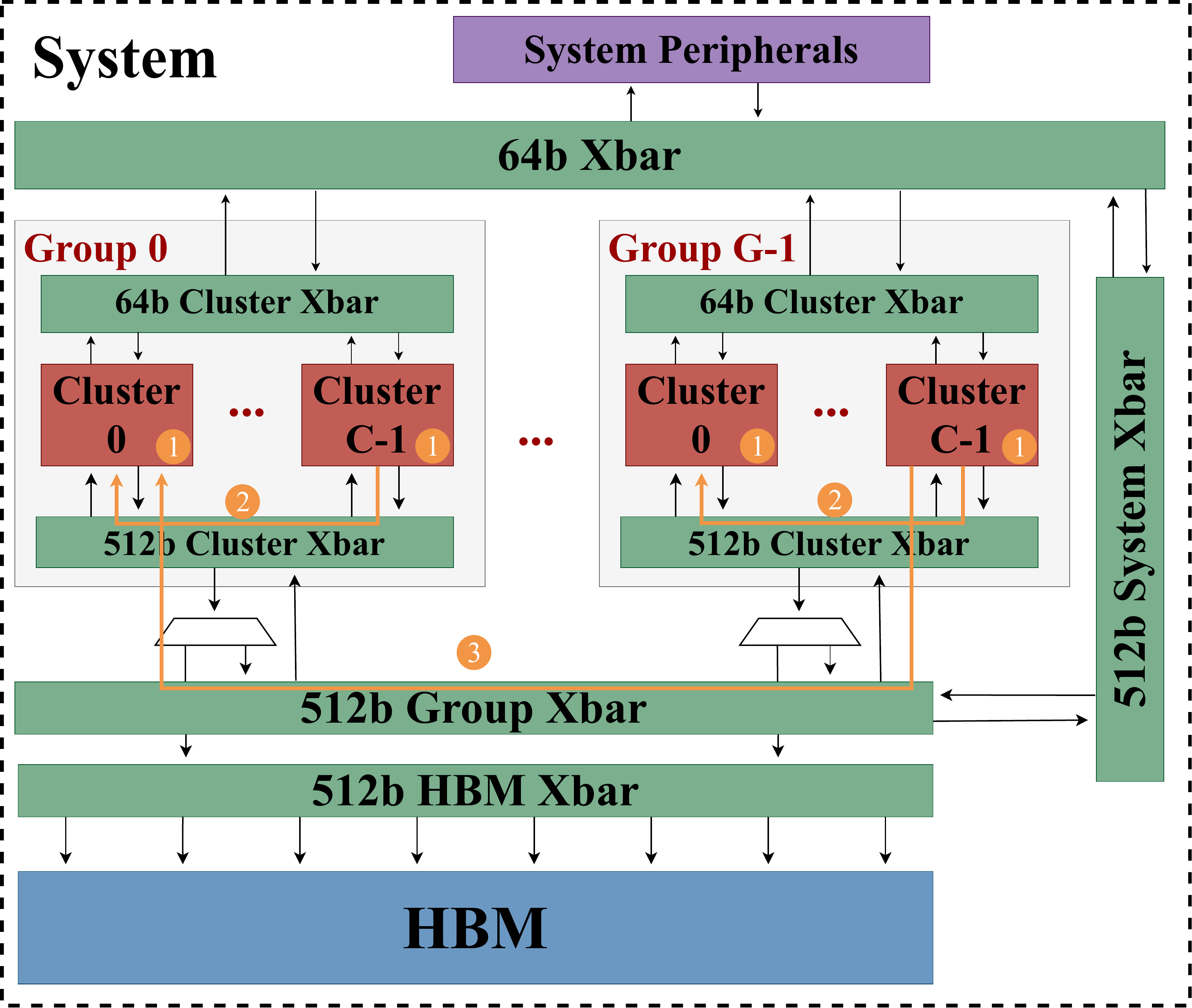}
    \caption{Scalable multi-cluster architecture with hierarchical heterogeneous memory interconnect.}
    \label{fig:memory}
\end{figure}

The compute cluster described in the previous section can be scaled up to a multi-cluster architecture as proposed in ~\cite{zaruba_manticore_2021} and silicon-proven in \cite{occamy_vlsi_2024}. A block diagram is shown in \autoref{fig:memory}. The first level of scalability is achieved with a \textit{Group} of $C$ compute clusters. The clusters of a group are connected via a narrow 64-bit crossbar for fast synchronization and a wide 512-bit \gls{axi} crossbar for efficient and high-bandwidth inter-cluster access.
The next level of scale-up is achieved by connecting $G$ Groups. Similar to the cluster level, a group-level \gls{axi} crossbar allows fast access between different groups. Additionally, the groups access eight \gls{hbm} channels over a wide \gls{hbm} crossbar for high-bandwidth access to the main memory.

This type of memory hierarchy enables increasingly higher bandwidth with each level while retaining scalability (see \autoref{fig:memory}). 
The first level, the cluster-to-\gls{spm} interconnect \numberincircle{1}, has a peak bandwidth of \SI{256}{\giga\byte\per\second}. The all-to-all \textit{Cluster} and \textit{Group} crossbars have a $C \times \SI{64}{\giga\byte\per\second}$ and $G \times \SI{64}{\giga\byte\per\second}$  peak bandwidth for inter-cluster \numberincircle{2} and inter-group \numberincircle{3} communication, respectively. The last level, the groups-to-\gls{hbm} connection, has a bandwidth of \SI{410}{\giga\byte\per\second}~\cite{noauthor_hbm2e_nodate} that can be fully exploited for configurations where $G$ is at least equal to the number of \gls{hbm} channels.
In our work, we design our \gls{fm} library to exploit each memory level and the heterogenous interconnect maximally: we primarily maximize the accesses to the local \gls{spm} memory. Then, we rely on cluster-to-cluster communication to efficiently transfer the data among clusters without intermediate copies to HBM. The access to the \gls{hbm} is limited to few accesses for loading and storing the input and output tensors, whereas intermediate tensors are stored in cluster memories.

%% file: Sections/05_fm_library.tex
\begin{figure*}[t]
    \centering
    \includegraphics[width=0.99\linewidth]{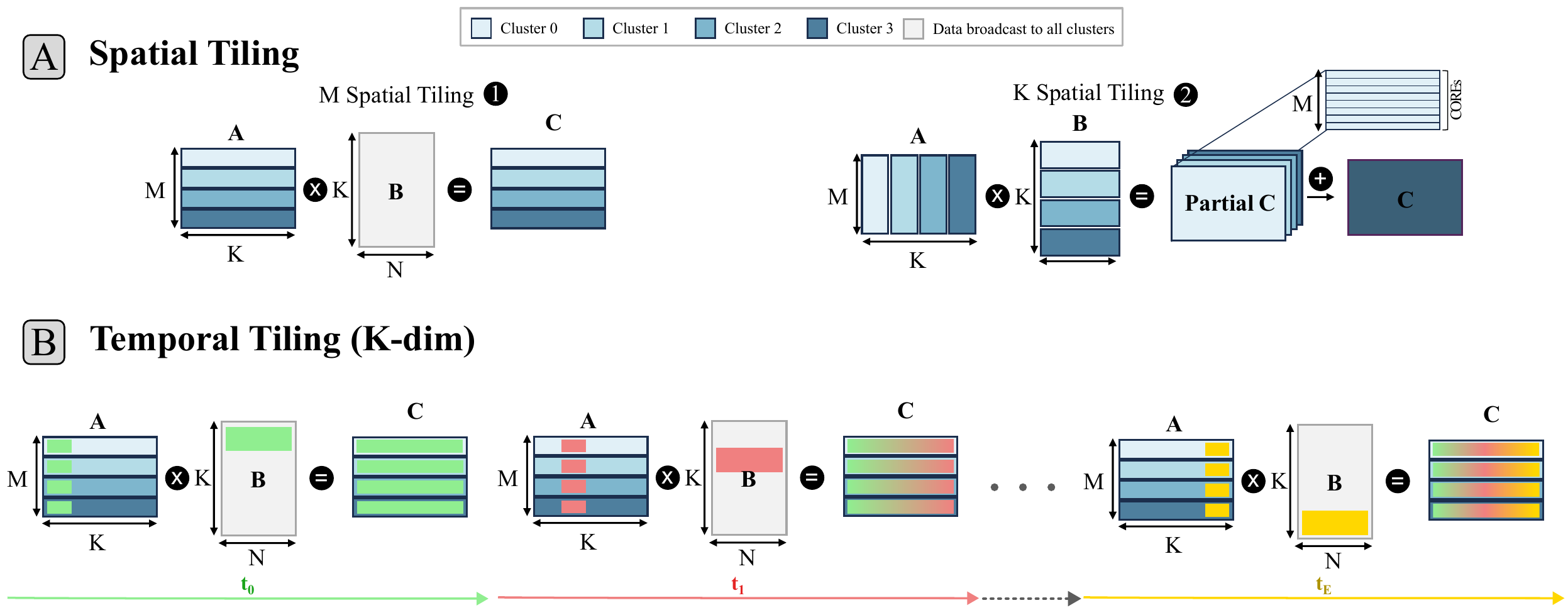}
    \caption{Illustration of the spatio-temporal GEMM Tiling. A) Spatial tiling of the \gls{gemm} operation on the $M$ and $K$ dimensions. When tiling $M$, each cluster processes distinct rows of the output matrix by partitioning matrix $A$ and $C$ blocks while broadcasting matrix $B$. When tiling $K$, matrices $A$ and $B$ are partitioned, while partial C matrices to be further summed together are produced for each cluster. B) Spatial tiling on the $M$ dimension combined with temporal tiling on the $K$ dimension where $t_0$, $t_1$, ..., $t_E$ denote different time steps. Note that at each individual time step, only a single temporal tile (in the figure, green, red, and yellow rectangles) is loaded in the cluster memory.}
    \label{fig:gemm}
    \vspace{-0.6cm}
\end{figure*}
\section{Foundation Model Library}%
\label{sec:library}

This section introduces our software stack to deploy Transformer-based \glspl{fm} efficiently onto the platform introduced in the previous section.
First, we describe our single-layer optimizations on the \gls{gemm}, Layernorm, \gls{gelu} kernels and our \gls{mha} layer implementation inspired by \cite{dao_flashattention-2_2023}. Then, we explain how we exploit layer fusion and the hierarchical interconnect of our hardware to maximize data reuse and minimize transfers from and to the HBM. Our \gls{fm} library supports \texttt{FP64}, \texttt{FP32}, \texttt{FP16}, and \texttt{FP8} layer variants and is available open-source  at \url{https://github.com/pulp-platform/snitch_cluster/tree/main/sw}. 

\subsection{Layer optimization}
\subsubsection{GEMM}
\label{sec:gemm}

In Transformer-based models, \gls{gemm} operations are fundamental components of the \gls{mha}, projection layers, and MLP layers. The GEMM operation computes $\mathbf{C} = \alpha \mathbf{A} \times \mathbf{B}$, where $\mathbf{A}$, $\mathbf{B}$, and $\mathbf{C}$ are matrices of size $M \times K$, $K \times N$ and $M \times N$, respectively, and $\alpha$ is an optionally present scalar multiplying factor.
We optimize three crucial knobs in our implementation: the spatial and temporal tiling (i.e., the partitioning of the layer computation between clusters and between successive iterations of the same cluster), the intra-cluster parallelization, and the \gls{gemm}'s innermost loop.

Matrix tiling is a key component of our implementation strategy.
Tiles are first spatially distributed between clusters and then temporally distributed over iterations. As shown in \autoref{fig:gemm}, we support spatial tiling in the $M$ and $K$ dimensions (\autoref{fig:gemm}-A) and temporal tiling in all matrix dimensions (an example of $K$-tiling is shown in \autoref{fig:gemm}-B). This condition is necessary to support all kernel sizes in our target \glspl{fm}, where spatial and temporal tiling along one or two dimensions alone would not suffice to fit tiles in the cluster-local L1 SPMs. Spatial tiling in the $M$ dimension maximizes utilization while avoiding frequent communication and synchronization among clusters. As shown in \autoref{fig:gemm}-B), temporal tiles of matrix $\mathbf{B}$ are broadcasted to all clusters while different blocks of matrixes $\mathbf{A}$ and $\mathbf{C}$ are distributed among the clusters. At each time-step i, we only load a single temporal tile of the matrices $\mathbf{A}$ and $\mathbf{C}$ in the cluster memories. We spatially tile on dimension $M$ instead of $N$ since the data are stored contiguously in memory with an $MN$ layout, and $M$ (which usually corresponds to the sequence dimension for \glspl{fm}) is always large enough for tiling.
\autoref{fig:gemm}-B shows the process over time: at each time step, a different portion of each spatial tile is loaded in each cluster memory, and a partial $C$ matrix result is produced and summed together with the previous ones generated. At the last time step $t_E$, the last partial $C$ spatial tiles are generated and summed with the previous ones (e.g., green plus red partial results in step $t_1$). The result of the \gls{gemm} is the concatenation of the spatial $C$ tiles from the different clusters.

Spatially tiling the $K$ dimension is generally not optimal since it generates partial sums in each cluster that need to be reduced. However, K-tiling becomes useful when the required input tiles of matrices $\mathbf{A}$ and $\mathbf{B}$ are already in the corresponding clusters because they have been produced by a previous operation, thus saving expensive memory transfers. Even if not reported in the Figure for the sake of space, again, only single temporal tiles are loaded in the cluster memory at each time step.
The \gls{mha} layer is the most notable example of this case. As the first \gls{gemm} operations ($Softmax(\nicefrac{QK^T}{\sqrt{P}})V$) are split over the heads (see Sec. \ref{sec:flashattention}), the following linear layer can be tiled on the heads dimension, corresponding to the internal $K$ dimension in our generic notation.
When $K$ is spatially tiled, the clusters' locally computed partial results are aggregated through a tree-wise reduction process, described in Sec.~\ref{sec:memory_access_optimization}.

Once each cluster is assigned a spatial-temporal tile, as described above, the computation is further parallelized at intra-cluster granularity.
Namely, the computation of each matrix tile is parallelized over the $M$ dimension by distributing distinct rows of the output matrix to different cores of the cluster (see \autoref{fig:gemm}).

At the level of every single core, the \gls{gemm}'s loop execution is optimized using the custom \gls{isa} extensions of the Snitch processor, specifically \texttt{Xfrep} and \texttt{Xssr}. We map the innermost loop $K$ to the FREP instruction to eliminate the indexing and branching overhead. 
The innermost loop computes the dot product between a row of the $\mathbf{A}$ matrix tile and a column of the $\mathbf{B}$ matrix tile. We map the $\mathbf{A}$ and $\mathbf{B}$ tiles to \glspl{ssr} 0 and 1. In this way, we can concurrently stream all operands to the \gls{fpu}. Furthermore, we unroll the innermost loop by a factor of 8 to hide the \gls{raw} stalls due to the \gls{fpu} pipeline latency.
Our single- and low-precision GEMM kernels are designed to exploit the \gls{simd} capabilities of Snitch's 64-bit wide \gls{fpu} for further parallelization, increasing the throughput of the innermost loop. In addition, the \texttt{FP16} and \texttt{FP8} kernels take advantage of Snitch's custom expanding dot-product \gls{isa} extensions.

Data movement is double-buffered at the cluster level using the \gls{dma} engine. Data are copied from/to the
main memory or from/to the SPM memory of clusters.


\subsubsection{FlashAttention-2}
\label{sec:flashattention}
\begin{figure*}[t]
    \centering
    \includegraphics[width=0.99\linewidth]{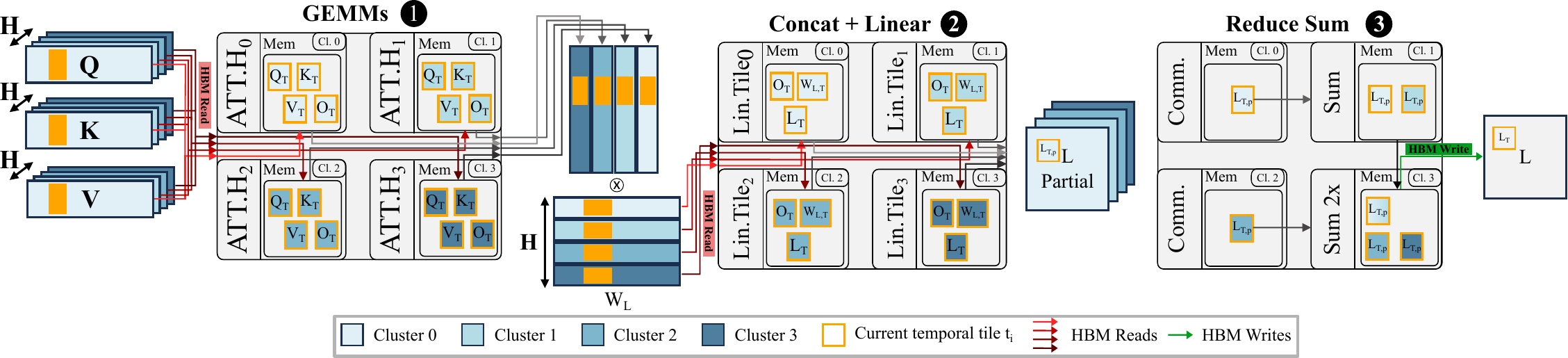}
    \caption{Example of \gls{mha} block mapping on an architecture with 4 clusters. Red/green arrows represent reads/writes from/to \gls{hbm}, and gray arrows represent intra- and inter-cluster data movement. The orange data chunks represent the data currently present in the clusters' \gls{spm} in a generic time step $t_i$.}
    \label{fig:mha}
    \vspace{-0.4cm}
\end{figure*}

The \gls{mha} block comprises two GEMM operations, a \softmax operation, and a final linear projection. We exclude the \gls{q}, \gls{k}, and \gls{v} projections from this paragraph since those are simple \gls{gemm} operations and can be optimized as discussed above.
To optimize \gls{mha}, we draw inspiration from the forward pass of the \emph{FlashAttention-2} algorithm proposed by Dao~\cite{dao_flashattention-2_2023}. This algorithm efficiently computes the fused scaled dot product attention by dynamically generating a tiled \softmax on the fly.
We implement the same dataflow proposed in the original FlashAttention-2 paper ~\cite{dao_flashattention-2_2023}.
We draw inspiration from the parallelization scheme presented therein, with the difference that \textit{we map attention heads to distinct Snitch clusters} (indicated with different blues in \autoref{fig:mha}) and that \textit{every cluster processes all the tiles} in the FlashAttention-2 dataflow in a time-iterative fashion in contrast to distributing tiles across GPU warps (in \autoref{fig:mha} we depict the execution of a single temporal tile at the generic time $t_i$, for each MHA sub-operation.)
All GEMMs involved in the individual attention head computations are parallelized exclusively within the associated cluster along the $M$ dimension, as described in Section \ref{sec:gemm}.
Similarly, we parallelize the calculation of the row statistics for the online \softmax across the cores in a cluster.
As noted by Dao, this strategy is embarrassingly parallel, so all clusters operate independently without the need for inter-cluster synchronization and communication.
As anticipated in Section \ref{sec:gemm}, the final linear layer is parallelized on the heads (corresponding to the $K$ dimension), exploiting a logarithmic reduction to accumulate partial results. \autoref{fig:mha} depicts the complete \gls{mha} kernel mapping. In orange, we show the data loaded from the HBM to the memory of each cluster in a generic time step $t_i$. Notice that at each time step, only a portion of the $L$ partial matrices is generated and stored by each cluster; then, a reduction-sum is performed over all clusters to generate a single temporal tile of the final $L$ matrix.

The optimized \texttt{FP32}, \texttt{FP16}, and \texttt{FP8} FlashAttention-2 layer implementations leverage the respective optimized GEMM kernels to accelerate the computation. All other operations, particularly the non-linear operations involved in the \softmax calculation, are performed in 32-bit \gls{fp} arithmetic in all implementations to avoid losses in accuracy. Conversion operations from low- to single-precision format are inserted at the output of the $\mathbf{Q}\times\mathbf{K^T}$ GEMM operation and before the final GELU. Vice versa, single- to low-precision conversions are performed before the final $\mathbf{A}\times\mathbf{V}$ GEMM operation.

\subsubsection{Layernorm}
The Layernorm operation is spatially tiled on the output matrix row dimension. Then, if the individual tiles do not fit the clusters' L1 \gls{spm}, they are temporally tiled on the column dimension.
Within a cluster, the rows in the assigned block are normalized in parallel by the eight compute cores. For the accumulation operations over the width of the row, we leverage the \glspl{ssr} in combination with the \texttt{Xfrep} extension. 
Lower precision formats exploit the SIMD arithmetic capabilities to parallelize normalization operations further.

\subsubsection{GELU}
To avoid costly division operations and the computation of the $tanh$ function, we approximate the \gls{gelu} with the i-\gls{gelu} polynomial described by Kim \etal{}~\cite{ref:ibert_2021}, which allows us to retain identical accuracy in all tasks benchmarked in the original paper.

\subsection{Memory Accesses Optimization}
\label{sec:memory_access_optimization}
Besides optimizing individual kernels, efficiently implementing attention layers requires careful consideration of memory access. In particular, maximizing data reuse in the higher levels of the memory hierarchy is crucial to reducing bandwidth requirements of the external \gls{hbm} and avoiding main memory access, which carries a high cost in terms of energy and latency.

To pursue this goal, we apply \textit{Layer Fusion} techniques in the \gls{mlp} and \gls{mha} blocks. 
No intermediate buffers are copied back to the HBM memory.
More specifically, in the \gls{mlp}, we fuse the computation of the element-wise \emph{\gls{gelu}} activation function with the preceding \emph{Linear} layer. 
We fuse the \emph{FlashAttention-2} layer in the \gls{mha} block with the subsequent Concat and Linear layers (see \autoref{fig:mha}). 
As explained below, this requires an additional reduction operation, for which we take advantage of our platforms' hierarchical interconnect. The hierarchcal interconnect allows us to reduce the data's on-chip traveling distance, improving overall energy efficiency and latency, and decreasing contention by making better use of the available bandwidth.
As can be noted in Fig. \ref{fig:attention}, thanks to our layer fusion and our hierarchical interconnect, we can reduce the total reads from the HBM by up to 1.6$\times$ (removing red memory transfers) for the GPT-J model, stepping from \SI{624}{\mega\byte} down to \SI{384}{\mega\byte}.

The right part of \autoref{fig:mha} shows the fusion details for the \gls{mha}. As the output of the GEMM block \blackcircle{1}, every cluster produces a different output matrix per head. While in an un-fused implementation, these blocks would have to be written back to the main memory to be subsequently loaded again by the following Concat layer. Instead, we use the locally computed matrixes without going through the \gls{hbm}. To this end, the $\mathbf{W_L}$ weights of the final linear layer are loaded in row-wise tiles into the different clusters, layer's GEMM along the $H$ dimension, producing many partial output matrices $L_c$, each of dimensions $S\times P$ \blackcircle{2}. 

Aggregating these partial results in a traditional shared-memory system would unavoidably mandate going through the main memory or the first shared cache level. As a consequence, accesses from all clusters would be serialized. Instead, thanks to the cluster-to-cluster connections in our hierarchical interconnect, we can implement the reduction in parallel and take advantage of the inherent data locality of nearby clusters without sending data to \gls{hbm}.
This is achieved by accumulating in a logarithmic reduction fashion among the clusters \blackcircle{3}. The depth $d$ of the binary reduction tree is determined as $d = \log_2(C\cdot G)$, where $C \cdot G$ is the total number of clusters.
Based on the current level of the binary tree, we determine whether a cluster is active and whether it will send or receive a tile of the partial result matrix. The sending cluster's \gls{dma} engine performs the transfer. If the cluster is designated as a \emph{receiver}, it will perform the sum reduction of the two tiles and proceed to the next level in the tree. The reduction is first performed among clusters in a group \numberincircle{2} and then among clusters in different groups \numberincircle{3} (see \autoref{fig:memory}). Finally, the last cluster stores the fully reduced $L$ matrix in \gls{hbm}.

\subsubsection{Double buffering} 
We employ \emph{double buffering} in all kernels to minimize the memory transfer overhead. The kernel process initiates with a priming phase, where the initial data set is loaded into the cluster's shared L1 \gls{spm}. Subsequently, the kernel computation begins. Concurrently, the dedicated \gls{dma} core preloads the next data chunk into the L1, preparing it for the next computation cycle, either from the main memory or from one other cluster L1 memory. This method ensures a seamless transition between computation and data movement phases without idle time, as data unloading and new computation iterations occur simultaneously. Therefore, we hide the latency associated with data transfers in and out of the memory.

%% file: Sections/06_Results.tex
\section{Experimental Setup}
\subsection{Foundation Models benchmarked}
\input{tables/02_models}

We considered five models to analyze our comprehensive deployment flow encompassing the \gls{fm} software library and the scalable multi-core HW platform.
All model hyperparameters are reported in Table \ref{tab:models}.
\emph{Params} denote the number of weights of the model, and \emph{FF} is the number of neurons of the first linear layers of the \gls{mlp} block.
The first three models are variants of the encoder-only \gls{vit}\cite{dosovitskiy_image_2020}, characterized by a different number of repeated transformer blocks, a different number of heads ($H$), and different values for the embedding and projection dimensions ($E$ and $P$, respectively).
The backbone of all three models is a series of repeated transformer blocks that include an \gls{mhsa} followed by a feed-forward \gls{mlp}. The input image is first divided into fixed-size patches by a convolutional layer; then, the patches are fed to the transformer network.  
The output of the final transformer layer is passed through a linear classifier head that maps the learned features to class predictions.
For this model family, we consider the image/s as the crucial benchmarking metric, i.e., the number of classifications produced per second.
Note that one classification is produced per run of the model.

The other two models, GPT-J and GPT3-XL, represent decoder-only \glspl{llm}.
The two models are characterized by their large scale and capacity to handle a wide range of tasks, from generating human-like text to solving complex coding problems.
Differently from the \gls{vit} variants, for these models, we consider two different execution modes: first, we report the results in \textbf{non-autoregressive (NAR) mode}, i.e., the mode used during the \textit{prompt encoding}. In this mode, the model generates a new sequence with the same length as the input. Each token of the produced sequence attends only to previous tokens in the input sequence. In other words, to predict token $t_i$, tokens $t_0, ... t_{i-1}$ should be used, while tokens $t_i, ..., t_S$ should be ignored. 
In practice, this is obtained by adding \textit{causal masking} \cite{dao_flashattention-2_2023} to the \gls{mhsa} block to avoid leakage of information from future tokens into the prediction of previous ones.
$S$ tokens are produced simultaneously in this mode with a single network execution\footnote{The results of this mode apply identically to the training \textit{forward pass} (with batch size = 1), which involves the same type of computation. We use this operating mode to compare with SoA platforms.}. 
The second mode considered for decoder-only models is the \textbf{autoregressive (AR) mode}, adopted during \textit{generative inference} after the prompt encoding. In this case, only a single next token is produced for each network invocation, using the previous part of the generated sequence as input. The primary metric used to benchmark decoder-only models for both modes is the number of tokens produced per second (tokens/s).
For all models, we also use GFLOPS as a general comparison metric.

\subsection{Simulation setup}
The software kernels are compiled with a customized LLVM 12 toolchain for Snitch with -O3 optimizations.
All performance results are obtained by cycle-accurate \gls{rtl} simulations with \textsc{Questasim} 2022.3.
We set the Snitch cluster at a frequency of \SI{1}{\giga\hertz} in a \SI{12}{\nano\meter} technology.
All main memory accesses per channel take a roundtrip access time of \SI{88}{\nano\second}~\cite{scheffler_sparse_2023}.
On average, we achieve a bandwidth with a four-cluster per group configuration of \SI{56}{\byte\per\cycle} per cluster for both reads and writes. We measured a \gls{dma} transfer setup time of \SI{27}{\nano\second} from RTL simulation. This amounts to a total static data transfer overhead of \SI{115}{\nano\second} per transfer.

We performed power measurements on a silicon prototype of our architecture with 16 clusters \cite{occamy_vlsi_2024} and 16 GiB of HBM2E.

\section{Experimental Results}
\label{sec:results}
\subsection{Impact of Software Optimizations}
\begin{figure}
  \centering
  \includegraphics[width=\linewidth]{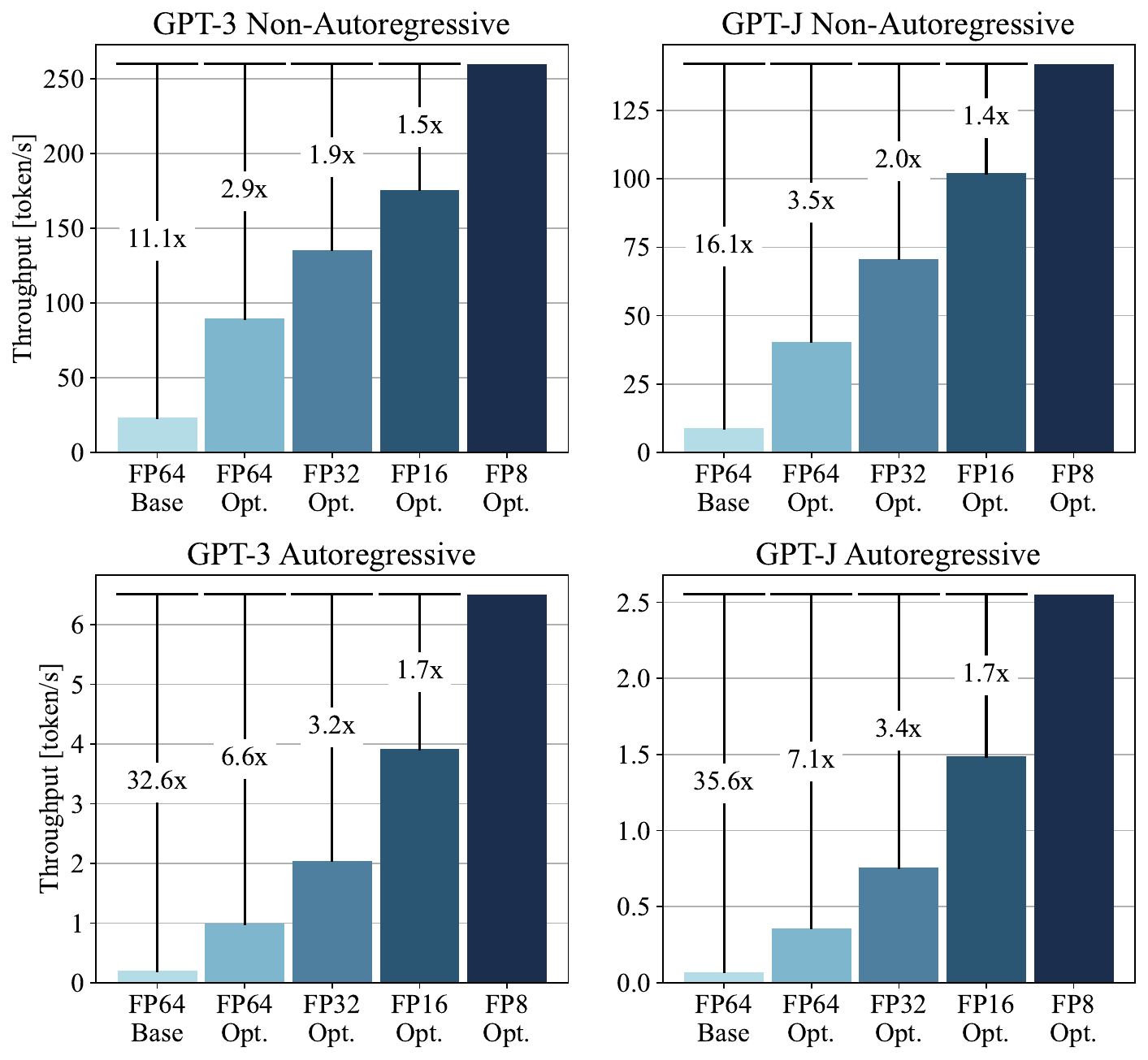}
  \caption{Impact of SW optimizations on the throughput of the \gls{gpt}-3XL and \gls{gpt}-J models with $S = 1024$ in the \gls{nar} and \gls{ar} modes.}
  \label{fig:gpt_sw_opt}
\end{figure}
\begin{figure}
  \centering
  \includegraphics[width=\linewidth]{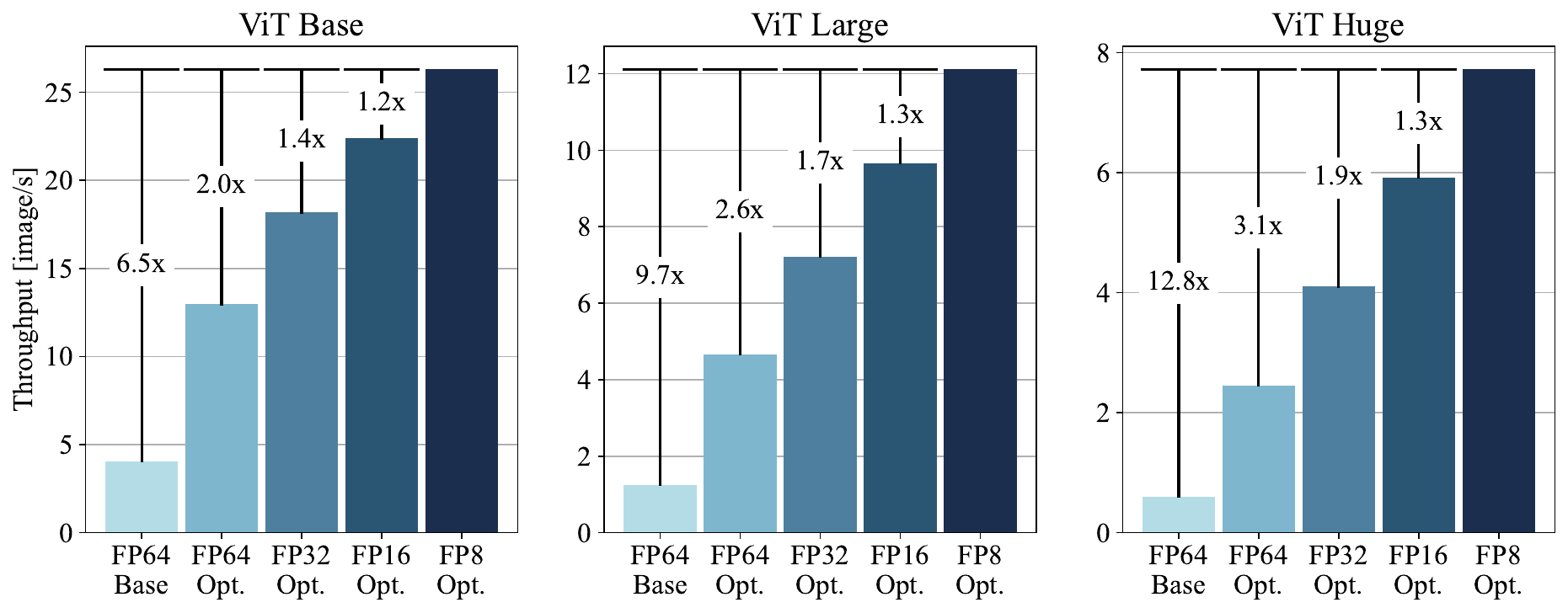}
  \caption{Impact of SW optimizations on the throughput of the \gls{vit} model class.}
  \label{fig:vit_sw_opt}
  \vspace{-0.4cm}
\end{figure}

In Fig. \ref{fig:gpt_sw_opt} and Fig. \ref{fig:vit_sw_opt}, we show the impact of the optimizations defined in Sec. \ref{sec:library}. 
We evaluate the throughput of the \gls{gpt}-3XL and \gls{gpt}-J in terms of generated output tokens per second on both the \gls{ar} and \gls{nar} operating modes. Furthermore, we investigate the impact of our optimizations on all three \gls{vit} models. 
The baseline implementation uses \fpd precision and does not exploit the \texttt{Xssr} and \texttt{Xfrep} extensions. Furthermore, the cluster-to-cluster communication of the hierarchical interconnect is not exploited.
By adding these features, we achieve an initial speedup of up to 5.0\x for the \gls{gpt} \gls{ar} mode, 4.6\x for the \gls{gpt} \gls{nar} mode, and 4.1\x for the \glspl{vit}. 
Then, on top of the \fpd implementation, we show the additional impact of lower-precision formats that exploit SIMD parallelism.
In particular, going to \fps, we observe an additional speedup of up to 1.8\x (\gls{nar}), 2.1\x (\gls{ar}) for the \gls{gpt}, and 1.6\x for \gls{vit} models, respectively. For the \gpt-J model, the improvement is even higher than the expected ideal one of 2\x due to the SIMD only. This improvement is possible thanks to the lower memory occupation of the \fps tensors, which allow for the fitting of larger tiles in the L1 \gls{spm}, leading to better parallelization and core utilization.
Going to 16-bit floating-point, we observe an additional speedup of 1.4\x (\gpt \gls{nar}), 2\x (\gpt \gls{ar}) and 1.5\x (\vit). 

With an \fpb data type, we achieve \textit{overall} speedups of up to 16.1\x for the \gls{gpt} models in \gls{nar} mode, 35.6\x in \gls{ar} mode, and 17.9\x for the \gls{vit} models. These correspond to 260/142 tokens/s in \gls{nar} mode and 6.5/2.6 tokens/s in \gls{ar} mode for GPT3-XL and GPT-J, respectively. For the three \gls{vit} models, we achieve 26, 12, and 8 image/s, respectively, for Base, Large and Huge models.

\subsection{Model \& HW Scalability}
\begin{figure}
  \centering
  \includegraphics[width=\linewidth]{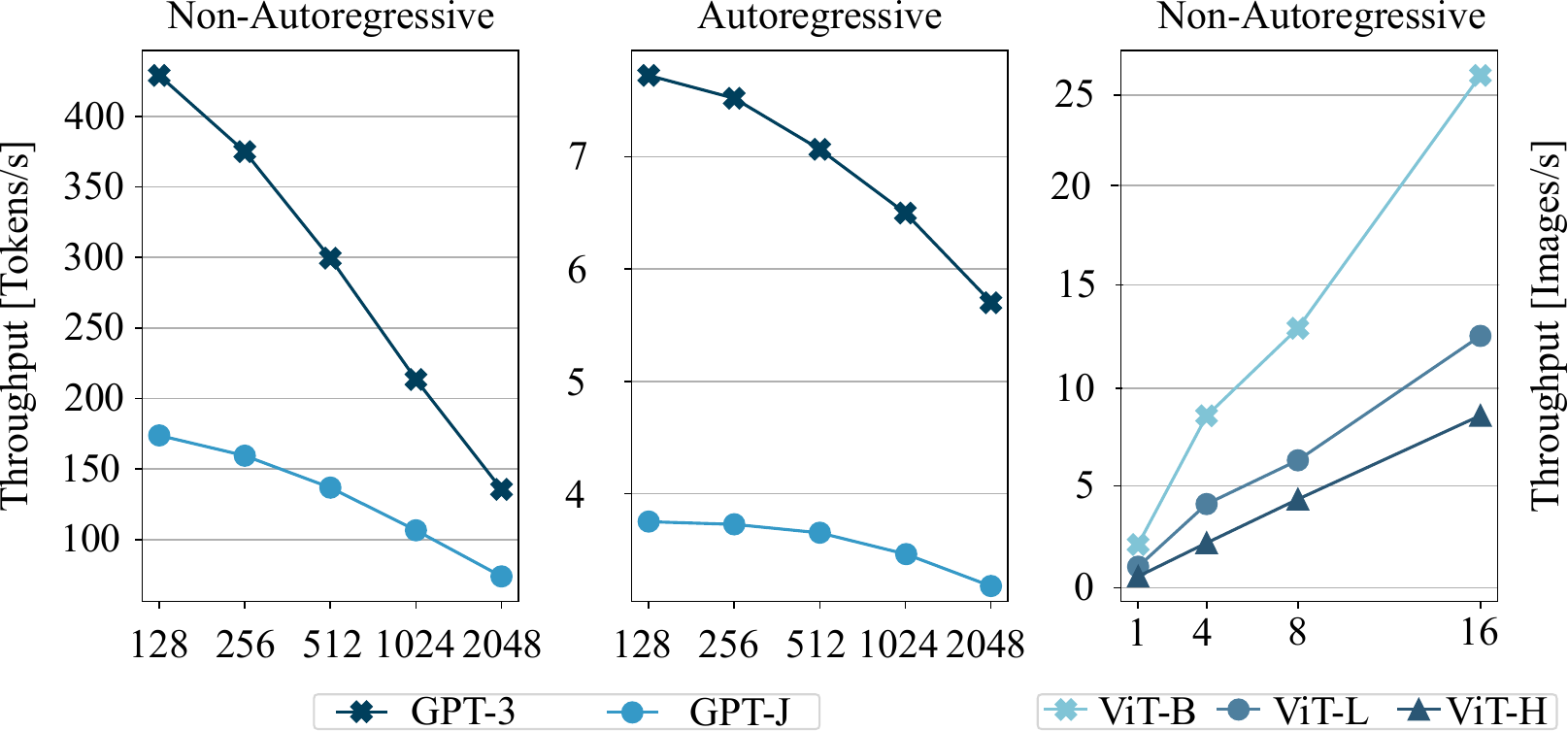}
  \caption{On the left, the impact of scaling the sequence length in the \gls{gpt} model class. On the right, scalability of the \gls{vit} models with increasing number of clusters.}
  \label{fig:scaling}
  \vspace{-0.3cm}
\end{figure}

Fig. \ref{fig:scaling} (leftmost two panes) shows the scaling of our end-to-end performance with respect to the input sequence length $S$. For this experiment, we consider GPT3-XL and GPT-J in \gls{ar} and \gls{nar} modes since \glspl{vit} are not normally fed with varying-length inputs.
In \gls{nar} mode, changing $S$ causes a quadratic increase in the computation complexity of the attention and a linear increase for the feed-forward layers. The number of tokens the network produces per invocation also varies (equal to $S$).
Overall, we observe that the end-to-end performance degrades with an almost constant slope proportional to the increase in operations, maintaining constant \si{\tflops}. Consequently, even when the sequence length increases, no overheads arise, such as expensive last-level memory spilling or lower tensor reuse.
Increasing the sequence length from 128 to 2048, the performance of GPT3-XL decreases from 429 tokens/s to 136 tokens/s, and GPT-J from 174 tokens/s to 74 tokens/s.

In \gls{ar} mode, the attention computation complexity grows linearly, while the linear layers' computation is independent of the sequence length, given that previous K and V elements are saved in the KV-cache. We observe this exact behavior when measuring individual operators' latencies.
Overall, for end-to-end \gls{ar} passes, performance ranges from 7.9 tokens/s to 5.8 tokens/s and from 3.8 tokens/s to 1 token/s for the GPT3-XL and the GPT-J models, respectively.

In both cases, \gls{gpt}3-XL exhibits a more pronounced increase in processing time per unit of sequence length, indicating that the computational load, and hence, resource utilization, can be distributed more effectively for larger models like \gls{gpt}-J.

Fig. \ref{fig:scaling} (rightmost pane) shows the images/s produced by the ViT models when increasing the number of clusters (and, therefore, of cores) in our scalable hardware platform.  When the number of clusters is lower than the heads, we apply temporal tiling to this dimension. When the number of clusters is identical or higher, we exploit this dimension for inter-cluster parallelization as described in \autoref{sec:library}. Compared to one cluster, we obtain a speedup of \{(4\x, 6\x, 12\x), (4\x, 6\x, 11.9\x), (4\x, 7.9\x, 15.8\x)\} for 4, 8, and 16 clusters on \vit-\{B, L, H\}, respectively. The progression of speedup nearly doubles with each step, which indicates a close-to-perfect scalability of our hardware and software stack. Namely, we are able to maintain similar hardware utilization despite the complexity of the additional inter-cluster communication.

\subsection{Kernel Latency Breakdown}
\begin{figure}
  \centering
  \includegraphics[width=\linewidth]{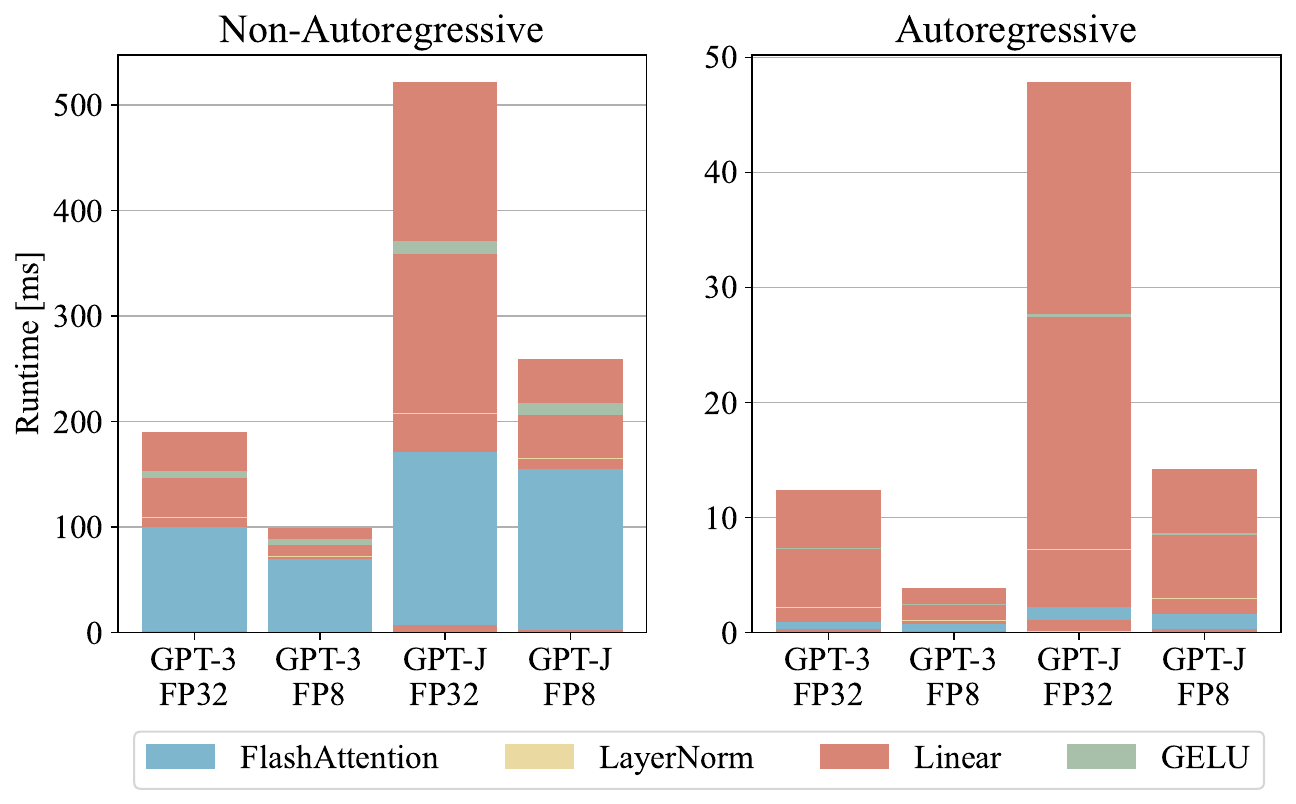}
  \caption{\texttt{FP32} and \texttt{FP8} kernel breakdown latency analysis in \gls{nar} and \gls{ar} modes.}
  \label{fig:gpt_latency_breakdown}
\end{figure}

\autoref{fig:gpt_latency_breakdown} shows the latency breakdown for the \gls{gpt}-J and GPT3-XL models in \fps and \fpb precision.
While the breakdowns differ for the two operating modes, the effect of precision reduction is identical.
Overall, we observe a significant latency reduction from \fps to \fpb, as also shown in Fig. \ref{fig:gpt_sw_opt}. This result can be analyzed in more depth by looking at the three main latency contributors.

First, we observe that activation layers, i.e., Layernorm and \gelu, have a limited latency impact, implying that they are not the primary bottlenecks within the model's architecture and that their scaling does not impact the overall layer latency.
Most latency is spent in GEMM operations – for GPT-J, 66\% in \fps and 36\% in \fpb of the total in \gls{nar} mode, 97\% in \fps, and 89\% in \fpb in \gls{ar} mode.
Conversely, the last latency contributor, the \fa kernel, has a more pronounced relative impact on the \fpb latency compared to \fps. This behavior is attributed to the \softmax exponential, still being executed in \fps, which does not fully capitalize on the latency reduction potential offered by \fpb. Furthermore, we need unpacking and packing operations moving from \fpb to \fps and vice-versa, which incurs additional latency in the \fpb version of the \fa kernel.
Executing the \softmax in \fps allows higher numerical stability and avoids quantization errors that could negatively impact the model's performance.
Exploring the execution of the \softmax at lower precision is out of the scope of our work.

\subsection{FPU Utilization \& Power Measurements}

\autoref{tab:power} summarizes the power, FPU Utilization, and \si{\gflops\per\watt} obtained by our \gls{fm} library, considering the different precisions on the GPT-J model with a sequence length of 1024 in both \gls{nar} and \gls{ar} modes. 

The first thing to notice is the utilization of the \gls{ar} mode. The \gls{ar} mode consistently achieves $<$ 10\% utilization of the FPU. This low utilization is caused by the fact that only a single token is processed while the previous ones are loaded from the KV cache. This strongly reduces the overall latency but at the cost of low utilization of the computational units.
On the other hand, when employing \gls{nar} mode, we consistently achieve utilization higher than 65\% when processing S output tokens in parallel. In particular, the wider the data type, the higher the utilization, given that the impact of non-arithmetic operations such as memory ones is lower. Furthermore, as previously said, in \fpb and \fph kernels, there are additional conversion operations to execute activation layers in \fps. 

In terms of power consumption, we observe a power proportional to the FPU utilization, with a max of \SI{5.2}{\watt} for the \fps \gls{nar} attention block.
In terms of efficiency, we reach the top performance of \SI{294}{\gflops\per\watt} with the \fpb kernel in \gls{nar} mode.
Nevertheless, even in the \gls{ar} mode, we can achieve the best performance of \SI{65.5}{\gflops\per\watt} thanks to the overall energy efficiency of the platform.

\input{tables/03_power}

\subsection{Comparison with State-of-the-Art}
\input{tables/04_soa_gflop_llm}
\autoref{tab:soa_gpt} compares our fully open deployment flow and platform against \gls{soa} commercial accelerators. We use the \gls{nar} mode of the \gls{gpt} class of models for this comparison to be compatible with SoA data. Emani \etal{}~\cite{emani_comprehensive_2023} collected data relative to the training forward pass (identical to our NAR mode) of GPT2-XL model running on many SoA accelerator platforms. All SoA platforms employ \fph. We also report our numbers in \fph to have a fair comparison. Nonetheless, we can achieve even higher throughput by utilizing \fpb precision. 
Given that commercial accelerators have a much larger scale compared to our platform, we do not benchmark only the total throughput but also the throughput divided by the number of compute units (CUs) and the FPU Utilization, computed as the ratio between the throughput achieved and the ideal maximum throughput of the platform.

Despite using general-purpose RISC-V cores with limited specialization for data-parallel compute patterns, our architecture achieves a throughput efficiency per compute unit comparable with the \gls{soa} of \SI{0.0056}{\tflops} employing \fph kernels. Only Gaudi2 and SN30 show impressively higher throughput; however, their CUs are either complex tensor processing units or matrix multiplication engines.
When comparing FPU Utilization, our architecture achieves the best result of 70.6\%, significantly outperforming \gls{soa} accelerators. For instance, the \emph{A100} and \emph{MI250} exhibit \gls{fpu} utilization of 14.42\% and 7.81\%, respectively. Even the \emph{Gaudi2}, which stands out among the compared platforms with an \gls{fpu} utilization of 34.62\%, reaches 2.04\x lower utilization than ours.

We also compare our results to those of the \nvidia's \emph{H100} \gls{gpu}. For this platform, the \emph{MLPerf} benchmarks~\cite{noauthor_benchmarking_2023} report a peak performance of 2683 samples/s on the (HF) \gls{vit}-large model in \fpb at 670W of power consumption, leading to an efficiency of 4 samples/s/W. Given the 17424 compute units (CUs) in H100, the throughput for CU corresponds to 0.15 samples/s/CU. In comparison, our architecture delivers a total of 27 samples/s for the same model in \fpb at a power consumption of \SI{4.5}{\watt}, corresponding to 0.2 samples/s/CU and 6 samples/s/W, outperforming the \emph{H100} by a factor of 1.3\x in terms of throughput per CU and 1.5\x in terms of efficiency.

On the other hand, given that academic accelerators run different workloads, such as sparse or dynamic transformers, and mostly use different data types, doing a direct comparison is not trivial. 
For instance, AccelTran performs inference only on the \emph{BERT-Tiny} architecture within a power envelope of \SI{14.03}{\watt} and 64 \glspl{pe}, i.e., \SI{0.22}{\watt}/\gls{pe}. In \gls{nar} mode, with a sequence length of $S=1024$, we operate in a power envelope of \SI{4.5}{\watt}, equating to \SI{0.04}{\watt}/\gls{pe}, i.e., 6.3\x more energy efficient.
The most comparable academic accelerator, developed by Tambe \etal{}, utilizes 8-bit and 4-bit \gls{mp} \emph{floating-point} computation for inference. They test their accelerator on the 12-layer 12-head BERT-base encoder model, achieving a minimum latency of \SI{489}{\milli\second} (normalized to 1 GHz to run at the same frequency of our architecture). We compared this number with our \fpb \vit-Base execution, considering that the models' parameters are almost identical (same $H$, Blocks, $E$, $FF$). On this model, we achieve \SI{38}{\milli\second} end-to-end inference, outperforming Tambe \etal's work by a factor of 12.8\x in terms of latency.

%% file: tables/02_models.tex
\begin{table}[t]
\caption{Foundation Models benchmarked.}
\label{tab:models}
\centering
\renewcommand\arraystretch{1.2}
\begin{tabular}{l|lll|ll}
 & ViT-B & ViT-L & ViT-H & GPT3-XL    & GPT-J      \\ \hline
Blocks                                                              &     12      &     24      &    32      &    40 & 28                    \\ \hline
Params &   86M       &    307M       &     632M     &     1.3B       &     6B       \\ \hline
E                                                                    & 768      & 1024      & 1280     & 2048       & 4096       \\ \hline
P                                                                    & 64       & 64        & 80       & 128        & 256        \\ \hline
S                                                                    & 197      & 197       & 197      & [128-2048] & [128-2048] \\ \hline
FF                                                                   &    3072      &    4096       &    5120      & 8192       & 16384      \\ \hline
H                                                                    & 12       & 16        & 16       & 16         & 16         \\ \hline
\end{tabular}
\end{table}

%% file: tables/03_power.tex
\begin{table}[ht]
\centering
\caption{Power measurements and Efficiency on NAR and AR workloads in different precisions.}
\label{tab:power}
\renewcommand\arraystretch{1.2}
\begin{tabular}{@{}clccc@{}}
\toprule
\textbf{Mode} &
  \multicolumn{1}{c}{\textbf{Implementation}} &
  \textbf{Power {[}W{]}} &
  \multicolumn{1}{l}{\textbf{GFLOPS/W}} &
  \textbf{FPU Util {[}\%{]}} \\ \midrule
                      & FP64 Optimized & 5.0 & 38.8  & 76.3  \\
                      & FP32 Optimized &  5.2 & 78.8  & 79.7  \\
                      & FP16 Optimized & 4.8 & 151 & 70.6 \\
\multirow{-4}{*}{NAR} & FP8 Optimized  & 4.5 & 294 & 65.2 \\ \midrule
\multicolumn{1}{l}{}  & FP64 Optimized & 2.1 & 10.0   & 8.32  \\
\multicolumn{1}{l}{}  & FP32 Optimized & 2.2 & 20.1  & 8.46  \\
\multicolumn{1}{l}{}  & FP16 Optimized & 2.1 & 38.3  & 7.89  \\
\multicolumn{1}{l}{\multirow{-4}{*}{AR}} &
  FP8 Optimized &
  {2.0} &
  65.6 &
  6.39 \\ \bottomrule
\end{tabular}
\end{table}

%% file: tables/04_soa_gflop_llm.tex
\begin{table}
\centering
\caption{Comparison with \gls{soa} accelerators of the \gls{gpt} \gls{nar} mode (GPT2-XL for SoA, GPT3-XL for our work) in \texttt{FP16}. CU: Compute Unit}
\label{tab:soa_gpt}
\renewcommand\arraystretch{1.2}
\begin{tabular}{@{}llllll@{}}
\hline  
        & \multirow{2}{*}{\textbf{A100}} & \multirow{2}{*}{\textbf{MI250}}  & \multirow{2}{*}{\textbf{SN30}} & \multirow{2}{*}{\textbf{Gaudi2}} & \multirow{2}{*}{\textbf{Ours}} \\
        &  &   &  &  &  \\
        \hline
        \textbf{Compute} & 6912 & 13312  & \multirow{2}{*}{1280} & 24 & \multirow{2}{*}{128} \\ 
        \textbf{Units} & +432 & +208 &  & +2 & \\ \hline
        \textbf{Throughput} & \multirow{2}{*}{5.63} & \multirow{2}{*}{3.75}  & \multirow{2}{*}{13.8} & \multirow{2}{*}{11.3} & \multirow{2}{*}{0.72} \\
         (TFLOPS)&  &  &  &  &    \\ \hline
        \textbf{Throughput} & \multirow{2}{*}{0.0008} & \multirow{2}{*}{0.0003}  & \multirow{2}{*}{0.0107} & \multirow{2}{*}{0.4327} & \multirow{2}{*}{0.0056} \\
         per CU (TFLOPS) &  &  &  &  &    \\ \hline
        \multirow{2}{*}{\textbf{FPU Util.} $[\%]$} & \multirow{2}{*}{14.4} & \multirow{2}{*}{7.8} & \multirow{2}{*}{16.0} & \multirow{2}{*}{34.6} & \multirow{2}{*}{\textbf{70.6}} \\
         &  &  &  &  &    \\
        \hline
    \end{tabular}
    \end{table}

%% file: Sections/07_Conclusions.tex
\section{Conclusion}
\label{sec:conclusion}
In this work, we systematically explored the deployment of \glspl{fm} on a multi-core scalable RISC-V platform. Our experimental analysis included three ViT models and two GPT models (GPT3-XL and GPT-J). We achieve superior FPU utilization compared to \gls{soa} commercial accelerators by tailoring our kernels to specific hardware resources, and memory interconnect characteristics.
In particular, by implementing specialized kernels for the attention mechanisms inherent in \glspl{fm}, we have shown speedups of over 35\x compared to the baseline implementation.
In our future work, we will explore additional SW and HW innovations to support \gls{fm} execution, such as multi-chiplet scalability. We aim to create a more adaptable, efficient, and powerful flow encompassing SW and HW components that can handle the increasing complexity and size of emerging \glspl{fm} in an open-source ecosystem.

\section*{Acknowledgments}
\noindent
This work has been supported in part by ‘The European Pilot’ project under grant agreement No 101034126 that receives funding from EuroHPC-JU as part of the EU Horizon 2020 research and innovation programme.